\documentclass[aps,pra,showpacs,twoside,twocolumn,10pt,floatfix]{revtex4-1}
\usepackage[colorlinks=true, citecolor=red, urlcolor=blue ]{hyperref}
\usepackage{epsfig,newlfont,amssymb,amsfonts,amsmath,bm,subfigure,graphicx,geometry,natbib,xcolor}
\usepackage{palatino,mathtools,amsthm,braket,soul,enumitem,color,algpseudocode}
\usepackage[normalem]{ulem}
\algrenewcommand\algorithmicindent{1.0em}

\definecolor{modified}{HTML}{3B5BB6}

\begin{document}
\title{
Uniform Decoherence Effect on Localizable Entanglement in Random  Multi-qubit Pure States
}
\author{Ratul Banerjee$^1$, Amit Kumar Pal$^{2,3}$, Aditi Sen(De)$^1$}
\affiliation{\(^1\)Harish-Chandra Research Institute and HBNI, Chhatnag Road, Jhunsi, Allahabad - 211019, India
\\
\(^2\)Faculty of Physics, University of Warsaw,  Pasteura 5, 02-093, Warsaw, Poland\\
\(^3\)Department of Physics, Indian Institute of Technology  Palakkad, Palakkad - 678557, India
}

\begin{abstract}

We investigate the patterns in distributions of localizable entanglement over a pair of qubits  for random multi-qubit pure states. We observe  that the mean  of localizable entanglement  increases gradually with increasing the number of qubits  of random pure  states while the standard deviation of the distribution decreases. The effects on the distributions, when  the random pure multi-qubit states are subjected to local as well as global noisy channels, are also investigated. Unlike the noiseless scenario,  the average value  of the localizable entanglement remains almost constant with the increase in the number of parties  for a fixed value of noise parameter. We also find out that the maximum strength of noise under which  entanglement  survives can be independent of the  localizable entanglement content of the initial random pure states.
\end{abstract}

\maketitle

\section{Introduction}
\label{sec:intro}

Multipartite entanglement~\cite{horodecki2009} -- one of the most important traits of composite quantum systems -- has been proven to be an useful ingredient in quantum protocols like  quantum teleportation~\cite{bennett1993,*bouwmeester1997,*murao1999,*grudka2004,*sende2010a}, quantum dense coding~\cite{bennett1992,*mattle1996,*sende2010,*bruss2004,*bruss2006,*horodecki2012,*shadman2012, *das2014,*das2015}, quantum cryptography~\cite{ekert1991,*jennewein2000,*gisin2002,*pirandola2019,*hillery1999,*cleve1999,*karlsson1999}, and measurement-based quantum computation~\cite{raussendorf2001,*hein2004,*hein2006, *briegel2009}. It also reveals interesting features in quantum many-body systems~\cite{amico2008,*chiara2018}, e.g., in quantum critical regions of the phase diagrams~\cite{orus2008} such as  spin chains~\cite{wei2005,*orus2010,*biswas2014},  in valence-bond solid states~\cite{orus2008b,*dhar2013}, and in topological systems~\cite{pezze2017}. This has motivated outstanding experimental advancements in creating multi-party entangled states in the laboratory using different substrates, such as atoms~\cite{*mandel2003,leibfried2005}, ions~\cite{monz2011,*friis2018}, photons~\cite{prevedel2009,*gao2010,*yao2012,*wang2018}, superconducting qubits~\cite{barends2014,*gong2019}, nuclear magnetic resonance molecules~\cite{negrevergne2006}, and very recently solid state systems~\cite{bradley2019}.  It has been shown that with increasing number of parties in the composite quantum systems, random pure states~\cite{bengtsson2006} tend to be  highly  multi-party entangled~\cite{hayden2006,*kendon2002,*enriquez2018,*klobus2019}, when entanglement  is quantified via distance-based measures~\cite{horodecki2009,shimony1995,*barnum2001,*wei2003}. While potential use of multipartite entanglement as resource in quantum protocols highlights the usefulness of this feature, it was shown that highly entangled random multi-party pure states may not be beneficial for computational speed-up~\cite{gross2009,*bremner2009} (cf.~\cite{jozsa2003}).  

A major roadblock towards the study of the characteristics and utility of multi-party entangled quantum states with higher number of parties is the limited availability of computable entanglement measures, both in pure as well as in  mixed states describing arbitrarily large composite quantum systems~\cite{horodecki2009}. This has motivated the search for quantum correlation measures that, on one hand, are computable for random multi-party quantum states, and on the other hand, exhibit properties similar to multi-party entanglement when the system-size is increased. Monogamy-based quantum correlations~\cite{coffman2003, dhar2017}, with measures belonging to both entanglement-separability~\cite{horodecki2009} and information-theoretic~\cite{modi2012,*bera2017} paradigms,  have recently been shown to exhibit properties similar to  multi-party entanglement for  random multi-party pure states~\cite{rethinasamy2019}. 

In this paper, we focus on the entanglement concentrated over  chosen subsystem(s) of the multi-party  \emph{random} quantum states via local independent projective measurements on the rest of the system, which is referred to as \emph{localizable entanglement} (LE)~\cite{verstraete2004,*verstraete2004a,*popp2005,*jin2004} (cf. entanglement of assistance~\cite{divincenzo1998,*smolin2005} and assisted mutual information~\cite{streltsov2015}). Such quantification and characterization of entanglement has been  shown to be appropriate and advantageous in graph states used in measurement-based quantum computation~\cite{hein2004,*hein2006}, in stabilizer states with and without noise~\cite{amaro2018,*amaro2019}, in defining correlation lengths in quantum many-body systems~\cite{verstraete2004,*verstraete2004a,*popp2005,*jin2004}, and in protocols like entanglement percolation in quantum networks~\cite{acin2007}. The concept of localizable entanglement has recently been generalized for concentrating entanglement over  multipartite  subsystems~\cite{sadhukhan2017}. It was argued that for a tripartite state, localizable entanglement is not a tripartite entanglement monotone  and can also not be considered as a bipartite measure~\cite{gour2006}. Therefore, the behavior of localizable entanglement for random pure states can not be inferred from the previous results based on multi-party distance-based and monogamy-based entanglement measures.

%In this paper, we concentrate on localizable entanglement accumulated over a bipartite subsystem of a multi-party quantum system. In this situation and for an appropriate choice of a computable entanglement measure over the chosen bipartite system, analytical computation of the quantity is possible in a number of scenarios. These include paradigmatic multi-party quantum states~\cite{sadhukhan2017} including the multi-qubit generalized GHZ~\cite{sadhukhan2017,greenberger1989,dur2000}, generalized W~\cite{sadhukhan2017,dur2000}, and the Dicke states~\cite{sadhukhan2017,dicke1954,kumar2017}, as well as quantum many-body systems with certain symmetries~\cite{venuti2005}.  

%Moreover, in the case of a tripartite quantum system, entanglement concentrated over a bipartite subsystem  is shown to be neither  a bipartite measure of entanglement nor a tripartite monotone~\cite{gour2006}, which highlights the importance of the characterization of entanglement localized over a bipartite subsystem in a composite quantum system. 
%In view of the trend of genuine multi-party entanglement in random pure states with a high number of parties, 

Towards this aim, we ask the following question: \emph{Do most of the  multi-party random pure states also possess high values of  localizable entanglement computed over a chosen bipartite subsystem?} We answer this query affirmatively. Specifically, we Haar uniformly generate three-, four-, and five-qubit random pure states and determine the LE in terms of entanglement of formation~\cite{wootters1998,wootters2001} over two of the qubits obtained via optimizing local projection measurement(s) on the rest. We observe that the values of localizable entanglement follow specific frequency distributions. To quantify the pattern, we determine different metrics of these distributions, such as  mean, standard deviation, and skewness, in order to understand how LE over a qubit-pair behaves on average  when the number of qubits in the system is increased. Our investigation indicates that the mean of  LE over a qubit-pair in the case of multi-qubit systems increases gradually with increasing the number of qubits. On the other hand, the standard deviation of the distribution of LE decreases when one increases the number of qubits from three to five. 

 To our knowledge, most of the studies on the  properties of random states is limited to pure states. In this paper,  we go beyond the pure states by performing a systematic study  of localizable entanglement when the initial random pure state is affected by local as well as global noise, and ask the question as to how the distribution of LE changes in presence of noise. This is a reasonable query in a laboratory set-up where creation of multi-party entangled states are always affected by certain decoherence. We show that the increasing trend in the average value of localizable entanglement for random states does not alter with the variation of qubits in presence of noise. From this perspective, we also study  the robustness of LE against different types of local and global noise considered in this paper. In particular, we evaluate the critical value of the strength of noise after which the localizable entanglement vanishes for a given randomly generated state and find that the amplitude damping noise destroys LE less than any other noisy channels for a fixed value of noise strength.

%To answer these two questions, 

%The number of qubits in the multi-qubit system is as per the computational feasibility of the values of localizable entanglement corresponding to an arbitrary quantum state in our numerical setup. 
% Moreover,  we consider paradigmatic models of local as well as global noise, and study the qualitative as well as quantitative changes in the frequency distributions of the values of localizable entanglement when these noise are locally applied on Haar uniformly generated three-, four-, and five-qubit states. 
 
We organize the rest of the paper as follows. In Sec.~\ref{sec:def}, we provide brief descriptions of localizable entanglement and the different models for noise considered in this paper. In Sec.~\ref{subsec:le_nonoise}, the frequency  distribution of the values of localizable entanglement in the case of Haar uniformly generated three-, four-, and five-qubit random pure states is discussed, and the corresponding frequency distribution metrics are calculated. Sec.~\ref{subsec:le_noise} describes the effect of local and global noise on these frequency  distributions. The robustness of localizable entanglement against different types of noise along with a comparative study of different noise models from this viewpoint is presented in Sec.~\ref{sec:robust}. The concluding remarks are in Sec.~\ref{sec:conclusion}.  

\section{Necessary Ingredients}
\label{sec:def}

In this section, we first give the definition of localizable entanglement for arbitrary multi-qubit states.  We also ponder on the different types of noises considered in this paper, and set the corresponding terminologies. We shall only deal with qubit systems in this paper, and the definitions as well as terminologies are tailored accordingly. 

\subsection{Localizable entanglement}
\label{subsec:le}

In a multi-qubit system constituted of $n$ qubits, the  maximum possible average entanglement that can be accumulated over a qubit pair by measuring independent local projection opeators on the rest of the $n-2$ qubits is called the localizable entanglement ~\cite{verstraete2004,popp2005,sadhukhan2017} over the pair of qubits. In this paper, we restrict ourselves to rank-$1$ projection measurements. Without any loss in generality, we denote the qubits in the $n$-qubit system by $1,2,\cdots,n$, among which the local projection measurements are performed over the $n-2$ qubits except the first two. For a quantum state $\rho_n$ describing an $n$-qubit system, the LE over the qubits $1$ and $2$ is given by 
\begin{eqnarray}
E_{12}(\rho_n)=\max \sum_{k=1}^{2^{n-2}}p_k\mathcal{E}(\varrho_{12}^{(k)}). 
\label{eq:le}
\end{eqnarray}
Here, the multi-index $k\equiv{k_{3}^\prime k_{4}^\prime\cdots k_{n}^\prime}$ denotes the outcome of the rank-$1$ projection measurements corresponding to the projector $P_{i}^{(k_i^\prime)}$ on the qubit $i=3,\cdots,n$, and $\mathcal{E}$ denotes a pre-decided bipartite entanglement  measure, also known as the \emph{seed measure}~\cite{sadhukhan2017}, which is  computed on the reduced post-measured state $\varrho_{12}^{(k)}$ of the qubit-pair, $(1,2)$. For two-qubit states, computable entanglement measures include entanglement of formation \cite{wootters1998}, which we will compute in this paper, and logarithmic negativity \cite{vidal2002}. The state  $\varrho_{12}^{(k)}$  reads as 
\begin{eqnarray}
\varrho_{12}^{(k)}=\text{Tr}_{3,4,\cdots,n}\left[\rho_n^{(k)}\right], 
\end{eqnarray}
where  $\rho_n^{(k)}$ is the post-measurement $n$-qubit state corresponding to the outcome $k$, given by 
\begin{eqnarray}
\rho_n^{(k)}=\frac{1}{p_k}\mathcal{M}_{k}\rho_n\mathcal{M}_k^\dagger.
\end{eqnarray}
The probability of obtaining the measurement outcome $k$ is $p_k=\text{Tr}\left[\mathcal{M}_{k}\rho_n\mathcal{M}_k^\dagger\right]$,
  where
\begin{eqnarray}
\mathcal{M}_k=I_1\otimes I_2 \otimes \left[\bigotimes_{i=3}^n P_i^{(k_i^\prime)}\right], 
\end{eqnarray}
with $I_1$ and $I_2$ being the identity operator in the Hilbert space of qubits $1$ and $2$. The maximization in Eq. (\ref{eq:le}) is performed over a complete set of local rank-$1$ projection measurements on the $n-2$ qubits, which, in general, is difficult to perform. In the case of qubit systems, the rank-$1$ projectors on a qubit $i$ can be parametrized  in terms of two real parameters $(\theta_i,\phi_i)$ as $P_{i}^{(k_{i}^\prime)}=\ket{k_i^\prime}\bra{k_i^\prime}$, $k_i^\prime=a,b$,  with~\cite{nielsen2010} 
\begin{eqnarray}
\ket{a}_i&=&\cos\frac{\theta_{i}}{2}\ket{0}_{i}+\text{e}^{\text{i}\phi_{i}}\sin\frac{\theta_{i}}{2}\ket{1}_{i},\nonumber\\
\ket{b}_i&=&\sin\frac{\theta_{i}}{2}\ket{0}_{i}-\text{e}^{\text{i}\phi_{i}}\cos\frac{\theta_{i}}{2}\ket{1}_{i},
\label{eq:parameters}
\end{eqnarray}
where $0\leq\theta_{i}<\pi$, $0\leq\phi_{i}\leq 2\pi$, $\{\ket{0}_{i},\ket{1}_{i}\}$ being the computational basis of the Hilbert space of qubit $i$. 

To answer the questions raised in this paper, one needs to compute the exact value of LE corresponding to arbitrary multi-qubit quantum states, pure or mixed,  with high number of qubits via performing the maximization involved in the definition of LE (Eq.~(\ref{eq:le})).  Although the parametrization in Eq.~(\ref{eq:parameters}) reduces the maximization to a $2(n-2)$ parameter optimization problem for an arbitrary $n$-qubit quantum state, obtaining the optimal basis for computing the exact value of LE can still be a challenging task when $2(n-2)$ is a large integer. This is due to  the fact that the optimization depends explicitly on the localizable entanglement function, which in turn depends on the seed measure $\mathcal{E}$. Except two-qubit states, 
exact computation of an entanglement measure for an arbitrary mixed quantum state is usually difficult~\cite{horodecki2009}.
 Besides, the determination of localizable entanglement also includes applications of the  measurement operators $\mathcal{M}_k$ corresponding to each  measurement outcome, which has a  dimensionality exponential in $n$, given by $2^{n-2}$.  
 %Also, reduction of the post-measured states, which are  represented by  density matrices having dimensionality $2^{n}$, is a difficult task, especially in noisy situations, where one has to deal with density matrices corresponding to mixed quantum states. 
 Also, reduction of the post-measured states, obtained via partial trace over  $2^{n-2}$ dimensions   is, in general, difficult, especially in noisy situations, where one has to deal with density matrices corresponding to mixed quantum states.   Due to these difficulties, exact values of localizable entanglement, and the corresponding optimal bases are known only in a handful of cases involving large number of qubits~\cite{verstraete2004,popp2005,sadhukhan2017,venuti2005,amaro2018,*amaro2019}, where certain properties of the quantum states under consideration are exploited. The present problem demands computation of the exact values of localizable entanglement over a pair of qubits in arbitrary $n$-qubit systems, for which analytical solution does not exist, and one has to consider numerical recipes. In this paper, we consider upto five-qubit states, each of which correspond to  a maximization of LE over $6$ real parameters. Considering the different challenges towards the exact computation  of  localizable entanglement, this is the maximum number of real parameters that can be handled with satisfactory numerical accuracy in our computational setup.

Note that the value and ease of computation of localizable entanglement depends also on the choice and computability of the seed measure $\mathcal{E}$ over the reduced state $\varrho_{12}^{(k)}$ of two qubits. In the situations where noise is applied to the system, one has to deal with a mixed state describing the $n$-qubit system, and the subsequent post-measured states $\varrho^{(k)}$ and reduced post-measured states $\varrho_{12}^{(k)}$ will also be mixed.  For the purpose of this paper, we consider entanglement of formation (EoF)~\cite{wootters1998,*wootters2001} as the chosen seed measure, which, for a generic two-qubit state $\varrho_{12}$, is defined as 
\begin{eqnarray}
\text{EoF}&=&-\frac{1+\sqrt{1-C^2}}{2}\log_2\left[\frac{1+\sqrt{1-C^2}}{2}\right]\nonumber \\ 
&&-\frac{1-\sqrt{1-C^2}}{2}\log_2\left[\frac{1-\sqrt{1-C^2}}{2}\right].
\end{eqnarray}
Here,  the concurrence, $C$, of the two-qubit system is given by 
\begin{eqnarray}
C=\max\left[0,\lambda_1-\lambda_2-\lambda_3-\lambda_4\right], 
\end{eqnarray}
with $\lambda_i$s $(\lambda_1\geq \lambda_2\geq\lambda_3\geq\lambda_4)$ being the eigenvalues of the Hermitian matrix $\varrho^\prime_{12}=\sqrt{\sqrt{\varrho_{12}}\tilde{\varrho}_{12}\sqrt{\varrho_{12}}}$, with $\tilde{\varrho}_{12}=\left(\sigma^y_1\otimes\sigma^y_2\right)\varrho_{12}^\star\left(\sigma^y_1\otimes\sigma^y_2\right)$.

\subsection{Noise Models}
\label{subsec:noise}

To analyze the consequence of decoherence in a multi-party domain, we consider  two different situations -- \textbf{Case 1.} local noise acting  identically on each individual qubits of an $N$ qubit state, and \textbf{Case 2.} a global noise acting on the entire system. As local noise, we consider single-qubit non-dissipative as well as dissipative noise models. Examples of the former include the phase-flip (PF) and the depolarizing (DP) noise channels, while the latter is represented by amplitude-damping (AD) noise. We employ the Kraus operator representation~\cite{nielsen2010,holevo2012} of the evolution $\rho_0\rightarrow\rho=\Lambda(\rho_0)$ of an initial single-qubit  state, $\rho_0$, under noise, where the operation $\Lambda(.)$ can be expressed by an operator-sum decomposition as 
\begin{eqnarray}
\rho = \Lambda(\rho_0)=\sum_{\alpha}K_\alpha\rho_0 K^\dagger_\alpha.
%= \sum_{\alpha}p_\alpha \tilde{K}_\alpha\rho_0 \tilde{K}^\dagger_\alpha,
\label{eq:evolve}
\end{eqnarray}
Here, $\{K_\alpha\}$ is the set of single-qubit Kraus operators satisfying $\sum_{\alpha}K_\alpha^\dagger K_\alpha=I$, with $I$ being the identity operator in the Hilbert space of a qubit. The single-qubit Kraus operators for the PF and DP channels can be represented by the Pauli matrices, $\sigma^i$ $(i=x,y,z)$, as
\begin{widetext}
\begin{eqnarray}
\text{Phase-flip noise}&:& K_0=\sqrt{1-\frac{p}{2}}I;\;K_1=\sqrt{\frac{p}{2}}\sigma^z,\nonumber \\
\text{Depolarizing noise}&:& K_0=\sqrt{1-\frac{3p}{4}}I;\;K_1=\sqrt{\frac{p}{4}}\sigma^x,\;K_2=\sqrt{\frac{p}{4}}\sigma^y,\;K_3=\sqrt{\frac{p}{4}}\sigma^z,\nonumber \\
\end{eqnarray}
\end{widetext} 
while for the AD channel, the Kraus operators are 
\begin{eqnarray}
K_0&=&\left(\begin{array}{cc}
   1 & 0 \\
   0 &  \sqrt{1-p}\\
  \end{array}\right),
  K_1=\left(\begin{array}{cc}
   0 & \sqrt{p} \\
   0 &  0\\
  \end{array}\right).
\end{eqnarray}
Here, $p$ ($0\leq p\leq 1$) can be interpreted as the strength of the noise in the channel. To study \textbf{Case 1},  the same type of single-qubit noise is applied on each of the $n$ qubits simultaneously and independently,  so that the evolution of the $n$-qubit system can also be represented  by an equation similar to Eq.~(\ref{eq:evolve}).  Mathematically, 
\begin{widetext}
\begin{eqnarray}
\rho_{n}^0\rightarrow \rho_{n}=\Lambda\left(\rho_{n}^0\right)=\sum_{\alpha}\left[K_{\alpha}^1\otimes K_\alpha^2\otimes \cdots\otimes K_{\alpha}^n\right]\rho_{n}^0\left[K_{\alpha}^{1^\dagger}\otimes K_\alpha^{2^\dagger}\otimes \cdots\otimes K_{\alpha}^{n^\dagger}\right],
\end{eqnarray}
\end{widetext}
where $\rho_{n}^{0}$ is the initial $n$-qubit state, and $\left\{K_\alpha^{i};i=1,2,\cdots,n\right\}$ is the set of single-qubit Kraus operators.

Apart from the local noise, we also consider the \emph{global} white noise of strength $p$ that takes an $n$-qubit  state $\rho_0$ to 
\begin{eqnarray}
\rho=\frac{(1-p)}{2^n}I+p\rho_0,
\label{eq:whitenoise}
\end{eqnarray}
where $I$ is the identity matrix in the Hilbert space of the $n$-qubit system. 

In the rest of the paper, we denote the $n$-qubit noisy state by  $\rho_n(p)$, which is evidently a function of $p$. The localizable entanglement, $E_{12}$, of $\rho_n(p)$ will, therefore, also  be a function of $p$. To keep the notation uncluttered, the LE  of $\rho_n(p)$ is referred as $E_{12}(n,p)$ for $p>0$, and as $E_{12}^0(n)$ when $p=0$ (i.e., for pure states). Note that $E_{12}(n,p)$ $\left(E_{12}^0(n)\right)$ can take different values for different states even with fixed $n$ and   $p$ (with fixed $n$).

%However, we remind ourselves that in the case of random multi-qubit states, $E_{12}(n,p)$ ($E_{12}^0(n)$) can take different values for different states even for fixed $n$ and $p$  (fixed $n$). 

\begin{figure*}
\includegraphics[width=0.75\textwidth]{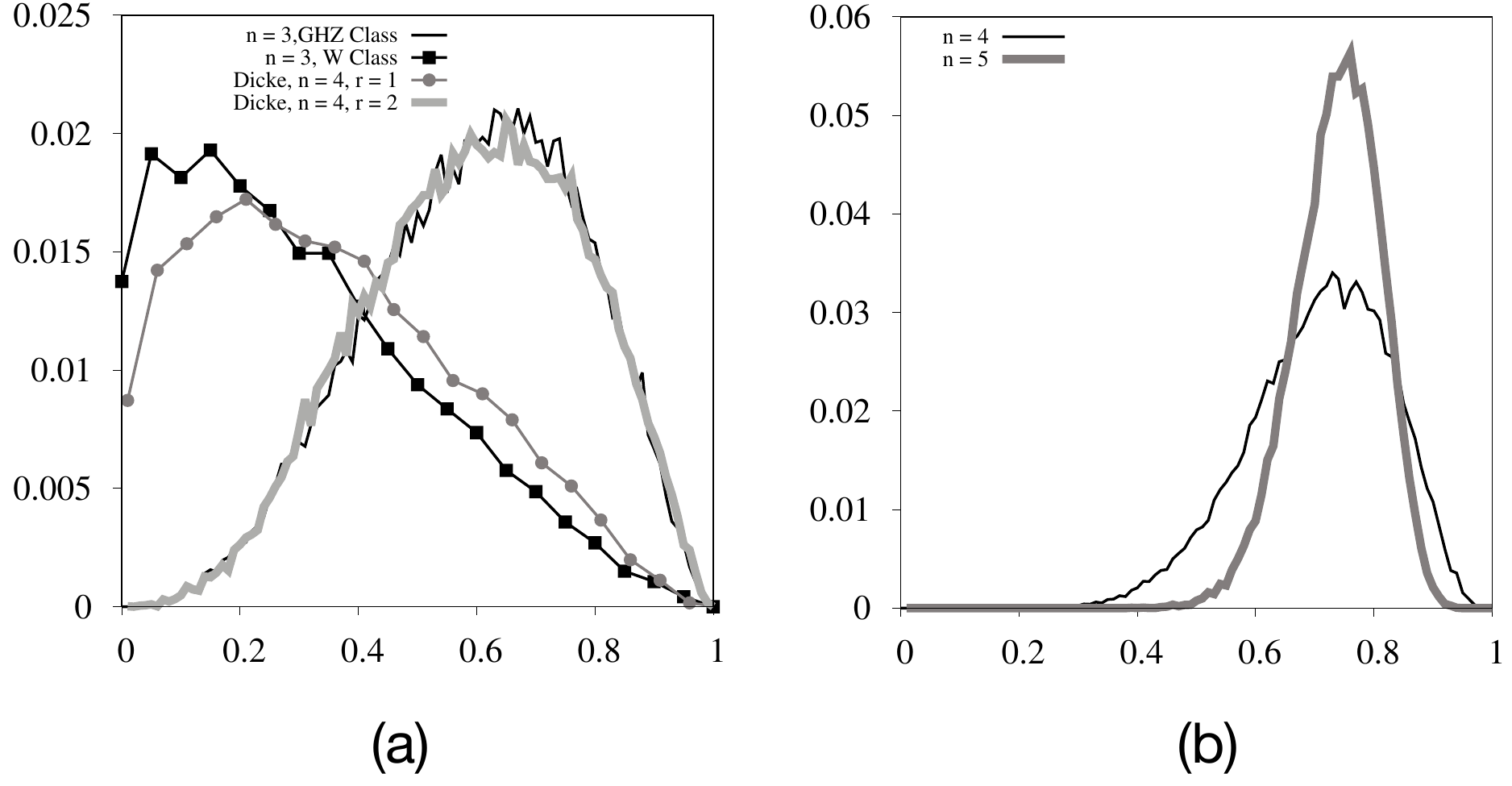}
\caption{(Color online.) \textbf{Absence of noise.} Normalized frequency distribution, $f_n^0$ (vertical axis), against $E_{12}^0(n)$ (horizontal axis). We Haar uniformly generate random pure states with (a) $n=3$, (b) $n=4,5$ for studying $f_n^0$ . We also simulate three-qubit W class (a) and $4$-qubit  generalized Dicke states with single and double excitations (a). The sample size is taken to be $5\times 10^{4}$ in each case. All quantities plotted are dimensionless.}
\label{fig:nonoise}
\end{figure*}

\section{Distribution of localizable entanglement}
\label{sec:multiparty}

As mentioned in Sec.~\ref{sec:intro}, random  pure states  with moderate values of $n$ are found to be  highly entangled~\cite{hayden2006,kendon2002} if one quantifies its entanglement via distance-based~\cite{horodecki2009,shimony1995,wei2003,barnum2001} or monogamy-based measures~\cite{rethinasamy2019}.  It is also noticed that  von Neumann entropy of local density matrices of random multi-party  pure states converges to unity for large number of parties~\cite{kendon2002}.  In this section, we address a similar question as to whether  random pure states shared between moderate number of parties  also possess high content  of  localizable entanglement. Such a question is non-trivial since LE is not straightforwardly a multi-party entanglement measure~\cite{gour2006}. Towards this aim, we first investigate the patterns of  frequency distributions of LE for random multi-qubit pure states and its variation with the increase in number of qubits. We then study the effects of noise on the distributions after sending all the qubits through noisy channels.  

\subsection{Noiseless scenario}
\label{subsec:le_nonoise}

%concentrated over the first two qubits of the $n$-qubit system in the noiseless scenario (i.e., $p=0$ in the case of different noisy channels defined in Sec.~\ref{subsec:noise}), which we denote by $E_{12}^0(n)$. We

Our aim here is to examine LE over first  two qubits of random pure  multipartite states with a chosen seed measure, specifically EoF~\cite{wootters1998,*wootters2001}. 
%Without any loss in generality, we concentrate on the first two qubits in the $n$-qubit system. 
A generic $n$-qubit pure state can be written as 
\begin{eqnarray}
\ket{\psi}=\sum_{j=0}^{2^n-1}a_j\ket{i_1}\ket{i_2}\cdots\ket{i_n} 
\label{eq:n_qubit_random_states}
\end{eqnarray}
with $\sum_{j=0}^{2^n-1}|a_j|^2=1$. Here, $\ket{i_k}\in\{\ket{0},\ket{1}\}$, $k=1,2,\cdots,n,$ form the computational basis of qubits $1$, $2$, $\cdots$, $n$. The state parameters $\{a_j;j=0,1,\cdots,2^n-1\}$ are complex numbers having the form $a_j=\alpha_j+\text{i}\beta_j$, $j=0,1,\cdots,2^n-1$, where $\alpha_j$ and $\beta_j$ are real numbers. For Haar uniformly generated $n$-qubit pure states, values of $\alpha_j$ and $\beta_j$, $j=0,1,\cdots,2^n-1$, can be  chosen from a Gaussian distribution of mean zero and standard deviation unity~\cite{bengtsson2006}. In  the   case of three-qubit systems, random pure states generated  in this fashion  belong to  the GHZ class of states~\cite{dur2000}.  We examine the normalized frequency distribution (NFD) of the values of localizable entanglement of formation,  which we obtain by Haar uniformly generating random pure states of $n=3,4,$ and $5$ qubits, and computing $E_{12}^0(n)$ for each of these states for a fixed value of $n$. The normalized frequency is defined as~\cite{bulmer1965} 
\begin{equation}
f_n^0=\frac{N({E_{12}^0(n)})}{N_S},
\label{eq:nfd_0}
\end{equation}
 where $N({E_{12}^0}(n))$ is the number of Haar uniformly generated random pure states of $n$ qubits having  $E_{12}^0(n)$, and $N_S$ is the total number of the $n$-qubit states simulated, representing the sample size. 
 
\small 
\begin{table}[ht]
\begin{tabular}{|c|c|c|c|c|}
\hline 
$n$ & 3 (W Class) & 3 (GHZ Class) & 4 & 5 \\ 
\hline 
$\left\langle E_{12}^{0}(n)\right\rangle$ &0.31  & $0.60$   & $0.71$ & $0.74$   \\
\hline
$\sigma_n$ & $0.22$  & $0.18$  & $0.12$  &  $0.07$   \\
\hline
$\eta_n$ & $0.61$  & $-0.28$ &$-0.46$  & $- 0.42$\\
\hline
\end{tabular}
\caption{\textbf{Noiseless scenario.} Table for the mean, standard deviation, and skewness of the normalized frequency distributions in the case of random three- (the GHZ class), four-, and five-qubit states and the W class states. All quantities are dimensionless.}
\label{tab:mean_nonoise}
\end{table}
\normalsize
 
\begin{figure*}
\includegraphics[width=0.9\textwidth]{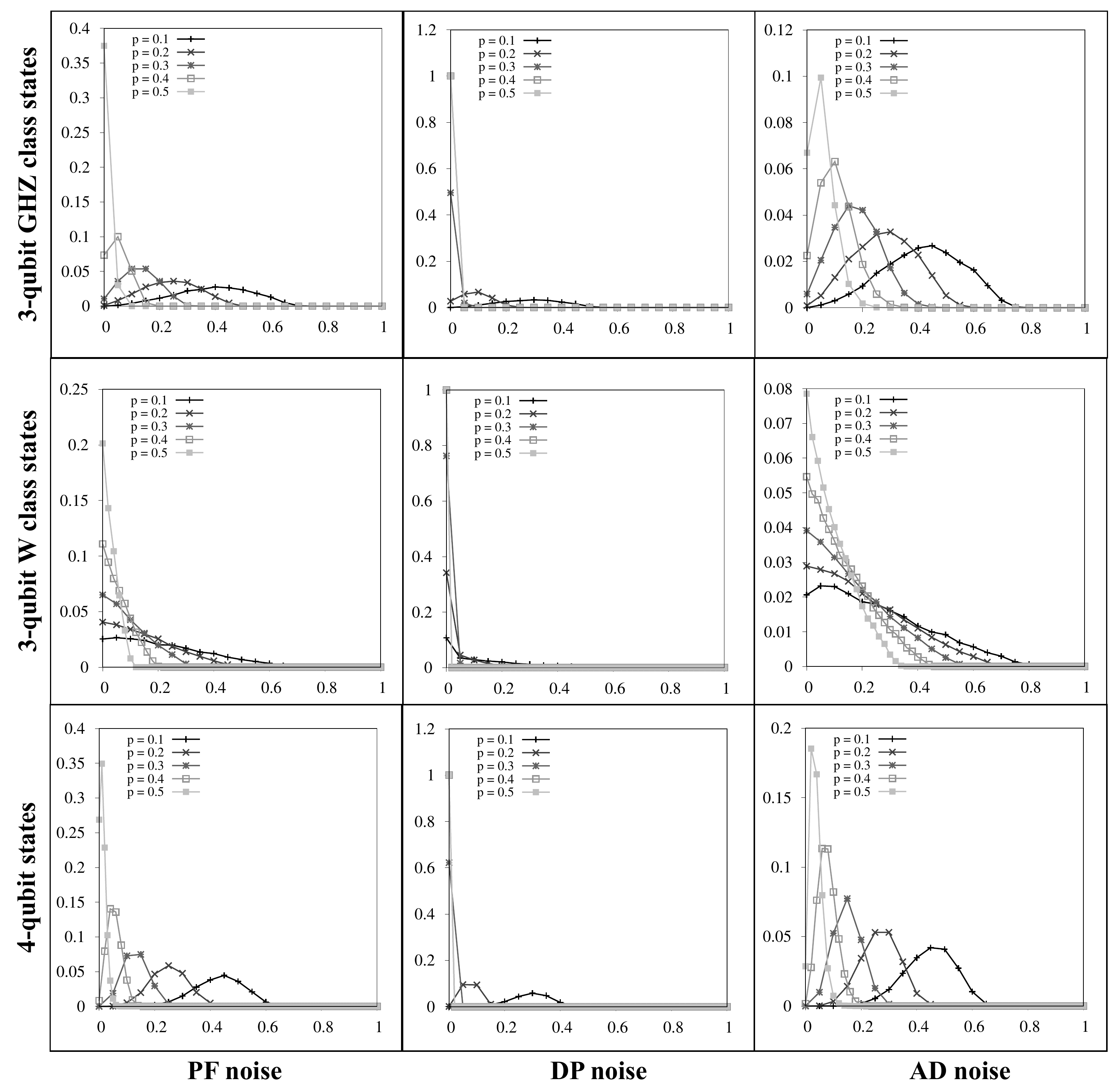}
\caption{(Color online.) \textbf{Normalized frequency distribution under local noise.} Normalized frequency distribution, $f_n^p$ (vertical axes), vs. $E_{12}(n,p)$ (horizontal axes). Random pure states in the case of $n=3$ (GHZ and W class) and $4$ (from top to bottom horizontal panels) are subjected to phase-flip (left column), depolarizing (middle column), and amplitude-damping (right column) noise of strengths $p=0.1,0.2,0.3,0.4,$ and $0.5$. A sample of $N_S=5\times 10^{4}$ random pure initial ($p=0$) states are considered for each of the normalized frequency distributions. All quantities plotted are dimensionless.}
\label{fig:noise}
\end{figure*}

The NFD of $E_{12}^0(n)$ in the case of three-, four-, and five-qubit systems are depicted in Fig.~\ref{fig:nonoise}. As is evident from the shape of the distributions, the mean of the NFD, given by~\cite{bulmer1965} 
\begin{eqnarray} 
\left\langle E_{12}^0(n)\right\rangle=\sum_{E_{12}^0(n)} E_{12}^0(n)f_n^0,
\label{eq:mean}
\end{eqnarray} 
shifts towards $\left\langle E_{12}^0(n)\right\rangle=1$ as $n$ increases from $3$ to $5$, thereby satisfying 
\begin{eqnarray}
E_{12}^0(n_1)> E_{12}^0(n_2)
\label{eq:compare_1}
\end{eqnarray}
for $n_1>n_2$ with $n_1,n_2\in\{3,4,5\}$. Note  that the increment is considerably lower in the case of the change $n=4$ to $n=5$ as compared to the increment during the change  $n=3$ to $n=4$. Interestingly however, the shapes of the distribution change drastically with the variation of $n$. To capture such feature of NFD, we  compute the standard deviation (SD)~\cite{bulmer1965}, 
\begin{eqnarray} 
\sigma_n=\left[\left\langle\left(E_{12}^0(n)\right)^2\right\rangle-\left(\left\langle E_{12}^0(n)\right\rangle\right)^2\right]^{\frac{1}{2}}. 
\label{eq:sd}
\end{eqnarray} 
Indeed, we find that the SD decreases remarkably as shown in Table~\ref{tab:mean_nonoise}, thereby showing more randomly generated states cluster around a large value of $\left\langle E_{12}^0\right\rangle$ with increasing  $n$. We also notice that with increasing $n$, the distributions become more symmetric around the mean, which we confirm by studying skewness~\cite{bulmer1965},  
\begin{eqnarray} 
\eta_n=\sum_{E_{12}^0(n)}\left(\frac{E_{12}^0(n)-\left\langle E_{12}^0(n)\right\rangle}{\sigma}\right)^3f(E_{12}^0(n)),
\label{eq:skew}
\end{eqnarray} 
of the NFD, tabulated in Table~\ref{tab:mean_nonoise}.

There exists another interesting class of three-qubit pure states, namely, the W class states~\cite{dur2000}, a generic state of which has the form 
\begin{eqnarray}
\ket{\psi}=a_0\ket{001}+a_1\ket{010}+a_2\ket{100}+a_3\ket{000}, 
\label{eq:3_qubit_w_class}
\end{eqnarray}
with $\sum_{l=0}^3|a_l|^2=1$. Similar to the case of the GHZ class states, the state parameters $\{a_j;j=0,1,2,3\}$ here are also complex numbers $a_j=\alpha_j+\text{i}\beta_j$, $j=0,1,2,3$, with $\alpha_j$ and $\beta_j$ being real numbers. Other states in W class either belongs to the subspace represented by $\ket{\psi}$, or are local unitarily connected to a state in that subspace.
% (see Fig.~\ref{fig:subspace} for a schematic representation).
From a three-qubit W class state having the form in Eq.~(\ref{eq:3_qubit_w_class}), a GHZ class state can not be obtained via stochastic local operations and classical communication (SLOCC) in a single-copy level~\cite{dur2000}. However, random W class states can be generated Haar  uniformly by  generating values of $\alpha_j$ and $\beta_j$, $j=0,1,2,3,$ from a Gaussian distribution of mean zero and standard deviation unity, which is a method similar to that for  the GHZ class states.

%\begin{figure} 
%\includegraphics[scale=0.5]{statespace.pdf}
%\caption{Schematic diagram representing the Haar uniformly generated subspaceof the W class (line AB), which is embedded in the space of 
%GHZ class states. The lines CD, EF, and GH represent the states in W class which are not generated, but are local unitarily connected to at least one state of the  generated sub-space.}
%\label{fig:subspace}
%\end{figure}

Let us justify that the generation of W class states in the procedure described above is Haar uniform. The state parameters $\{a_j;j=0,1,2,3\}$ are complex numbers $a_j=\alpha_j+\text{i}\beta_j$, $j=0,1,2,3$, where $\alpha_j$ and $\beta_j$ are real numbers.
% We generate random W class states by  choosing values of $\alpha_j$ and $\beta_j$, $j=0,1,2,3,$ from Gaussian distributions of mean zero and standard deviation unity. The justification behind this procedure is as follows. 
The parameters $\{\alpha_j,\beta_j\}$ form an eight variable tuple, which we denote by \(r = \{a_{j}, b_{j}\}\), where $r \in \mathbb{R}^{8}$, the real space in eight dimension. The individual probability distributions corresponding to $a_j$ and $b_j$ can be written as $\phi(a_{j}) = (2\pi)^{-1/2}\exp(-a^{2}_{j}/2)$ and $\phi(b_{j}) = (2\pi)^{-1/2}\exp(-b^{2}_{j}/2)$, respectively, and the joint probability density function of $r$ reads as $f(r) = (2\pi)^{-4}\exp\left[-\sum_{i=1}^{4} (a^{2}_{j} + b^{2}_{j})/2\right]$, which equals to  $(2\pi)^{-4}\exp\left[-\left \| r \right \|^{2}/2\right]$. Here, $f(r)$ is independent of the direction of $r$, but depends on the length of $\left \| r \right \|$, implying that $\hat r = \frac{r}{\left \| r \right \|}$ is uniformly distributed over $\mathbb{S}^{7}$, with $\mathbb{S}^7$ being seven-dimensional surface of unit sphere in $\mathbb{R}^{8}$, and we have considered normalization constraint in $\mathbb{R}^{8}$. It also implies that $\hat r$ follows the joint probability distribution $f(\hat r) = \exp(-1/2)/\sqrt{(2\pi)^8}$, which remains constant over all directions, thereby suggesting that the generated states are Haar uniform in this subspace (see~\cite{ozols2009}). In this paper, we consider localizable entanglement, which is invariant under local unitary transformation. Therefore, localizable entanglement of the Haar uniformly generated states of the form $\ket{\psi}$ in the generated subspace contains full  information about the behaviour of LE for the complete state space of the entire W class. This also implies that the statistics corresponding to the localizable entanglement in the generated subspace faithfully represents the statistics of it in the entire state space of the W class.

The W class states with vanishing monogamy score for concurrence form a set of measure zero in the states space of three-qubits~\cite{dur2000},   and hence they can not be found from the generation of three-qubit random states following the methodology descried in the discussion succeding Eq.~(\ref{eq:n_qubit_random_states}). This can easily be seen from  the fact  that the generation  of a generic state in  the W  class requires vanishing of a number of coefficients in the general form of a state in the GHZ class, which is not possible in the random Haar uniform generation~\cite{bengtsson2006} of GHZ class states having the form given in Eq.~(\ref{eq:n_qubit_random_states}). However, states belonging to W class also possess certain features of entanglement, which make them useful for quantum information processing tasks~\cite{sende2003, *kaszlikowski2008,*barnea2015,*laskowski2015, *roy2018}. We, therefore, separately generate  the W class  states by randomly choosing its parameters following the process described above, and determine the NFD of $E_{12}^0(n)$ (see  Fig.~\ref{fig:nonoise}) for comparison.

\small 
\begin{table*}[ht]
\begin{tabular}{c}
Depolarizing noise\\
\begin{tabular}{|c|c|c|}
\hline
$p=0.1$ & $p=0.2$ & $p=0.3$  \\
\hline
          \begin{tabular}{c|c|c|c}
            $n$ & $3$ & $4$ & $5$ \\ 
          \hline 
          $\left\langle E_{12}(n,p)\right\rangle$ & $0.30$  &$0.30$  &  $0.27$ \\
          \hline
         $\sigma_n(p)$ & $0.11$  &$0.07$  &$0.04$   \\
         \hline
         $\eta_n(p)$ & $-0.10$  & $-0.28$ &$-0.16$   \\
         \end{tabular}         
         &         
          \begin{tabular}{c|c|c|c}
          $n$ & $3$ & $4$ & $5$ \\
          \hline 
          $\left\langle E_{12}(n,p)\right\rangle$ & $0.10$  & $0.08$  &   $0.06$\\
          \hline
         $\sigma_n(p)$ & $0.05$  &$0.03$  & $0.02$  \\
         \hline
         $\eta_n(p)$ & $0.29$  & $0.30$ & $0.36$  \\
         \end{tabular}             
         &         
          \begin{tabular}{c|c|c|c}
          $n$ & $3$ & $4$ & $5$ \\
          \hline 
          $\left\langle E_{12}(n,p)\right\rangle$ & $0.01$ & $0.01$ & $0.00$  \\
          \hline
          $\sigma_n(p)$ & $0.01$ &$0.01$  &$0.00$   \\
         \hline
        $\eta_n(p)$ &  $1.15$ & $1.08$ & $1.03$  \\
         \end{tabular}  \\       
%         &         
%          \begin{tabular}{c|c|c}
%          $3$ & $4$ & $5$ \\ 
%          \hline 
%          $0.00$  &$0.00$  &$0.00$   \\
%          \hline
%         $0.00$ & $0.00$ & $0.00$  \\
%         \hline
%         $0.00$  &$0.00$  & $0.00$  \\
%         \end{tabular}        
%         &        
%          \begin{tabular}{c|c|c}
%          $3$ & $4$ & $5$ \\
%          \hline 
%          $0.00$  &$0.00$  &$0.00$   \\
%          \hline
%         $0.00$ & $0.00$ & $0.00$  \\
%         \hline
%         $0.00$  &$0.00$  & $0.00$  \\
%         \end{tabular} \\
\hline          
\end{tabular}\\
Phase-flip noise  \\
\begin{tabular}{|c|c|c|}
\hline 
$p=0.1$ & $p=0.2$ & $p=0.3$  \\
\hline 
          \begin{tabular}{c|c|c|c}
          $n$ & 3 & 4 & 5 \\ 
          \hline 
          $\left\langle E_{12}(n,p)\right\rangle$ & $0.40$   & $0.44$ & $0.43$  \\
          \hline
         $\sigma_n(p)$ & $0.14$  & $0.09$ & $0.06$  \\
         \hline
         $\eta_n(p)$ & $-0.19$   & $-0.28$ & $-0.11$  \\
         \end{tabular}         
         &         
          \begin{tabular}{c|c|c|c}
          $n$ & 3 & 4 & 5 \\ 
          \hline 
           $\left\langle E_{12}(n,p)\right\rangle$ & $0.24$  & $0.25$  &   $0.25$\\
          \hline
         $\sigma_n(p)$ & $0.10$  &$0.06$  &$0.05$   \\
         \hline
         $\eta_n(p)$ & $-0.04$  & $-0.02$ & $0.08$  \\
         \end{tabular}             
         &         
          \begin{tabular}{c|c|c|c}
          $n$ & 3 & 4 & 5 \\ 
          \hline 
          $\left\langle E_{12}(n,p)\right\rangle$ & $0.13$  & $0.13$ & $0.13$  \\
          \hline
        $\sigma_n(p)$ & $0.06$ & $0.04$ &$0.03$   \\
         \hline
        $\eta_n(p)$ & $0.13$  & $0.15$ & $0.16$  \\
         \end{tabular}      \\   
%         &         
%          \begin{tabular}{c|c|c}
%          3 & 4 & 5 \\ 
%          \hline 
%          $0.06$  & $0.06$  & $0.06$  \\
%          \hline
%         $0.04$  &$0.02$  & $0.02$  \\
%         \hline
%         $0.41$ & $0.35$ &  $0.25$ \\
%         \end{tabular}        
%         &        
%          \begin{tabular}{c|c|c}
%          3 & 4 & 5 \\ 
%          \hline 
%          $0.02$  &$0.02$  &  $0.02$ \\
%          \hline
%         $0.02$ &$0.01$  &$0.01$   \\
%         \hline
%         $1.10$ & $0.86$ & $0.52$  \\
%         \end{tabular}   \\
\hline                      
\end{tabular}\\
Amplitude-damping noise\\
\begin{tabular}{|c|c|c|} 
\hline 
$p=0.1$ & $p=0.2$ & $p=0.3$\\
\hline 
          \begin{tabular}{c|c|c|c}
          $n$ & 3 & 4 & 5 \\ 
          \hline 
          $\left\langle E_{12}(n,p)\right\rangle$ & $0.42$  & $0.45$ &$0.43$   \\
          \hline
         $\sigma_n(p)$ & $0.14$  &$0.09$  & $0.05$  \\
         \hline
         $\eta_n(p)$ & $-0.18$ & $-0.34$ &$-0.26$   \\
         \end{tabular}         
         &         
          \begin{tabular}{c|c|c|c}
          $n$ & 3 & 4 & 5 \\ 
          \hline 
         $\left\langle E_{12}(n,p)\right\rangle$ &  $0.28$  & $0.27$  & $0.24$  \\
          \hline
      $\sigma_n(p)$ &   $0.11$  & $0.07$ & $0.04$  \\
         \hline
        $\eta_n(p)$ &   $0.00$  & $-0.10$ &  $-0.01$ \\
         \end{tabular}             
         &         
          \begin{tabular}{c|c|c|c}
         $n$ &  3 & 4 & 5 \\ 
          \hline 
        $\left\langle E_{12}(n,p)\right\rangle$ &   $0.18$  & $0.15$ & $0.12$  \\
          \hline
         $\sigma_n(p)$ &   $0.08$  &$0.05$  &$0.03$   \\
         \hline
        $\eta_n(p)$ &  $0.23$ &$0.21$  & $0.28$  \\
         \end{tabular}         \\
%         &         
%          \begin{tabular}{c|c|c}
%          3 & 4 & 5 \\ 
%          \hline 
%          $0.11$ & $0.08$ &  $0.06$ \\
%          \hline
%         $0.06$  & $0.03$ & $0.02$  \\
%         \hline
%         $0.54$  & $0.55$ &  $0.58$ \\
%         \end{tabular}        
%         &        
%          \begin{tabular}{c|c|c}
%          3 & 4 & 5 \\ 
%          \hline 
%          $0.06$ & $0.04$ & $0.02$  \\
%          \hline
%         $0.04$ &$0.02$  & $0.01$  \\
%         \hline
%        $0.87$  & $0.92$ & $0.87$  \\
%         \end{tabular}     \\
\hline                                           
\end{tabular}\\

White noise\\
\begin{tabular}{|c|c|c|} 
\hline 
$p=0.1$ & $p=0.2$ & $p=0.3$ \\
\hline 
          \begin{tabular}{c|c|c|c}
          $n$ & 3 & 4 & 5 \\ 
          \hline 
          $\left\langle E_{12}(n,p)\right\rangle$ & $0.46$  & $0.55$ & $0.57$  \\
          \hline
         $\sigma_n(p)$ & $0.15$  & $0.10$ & $0.06$  \\
         \hline
         $\eta_n(p)$ & $-0.26$ & $-0.45$ & $-0.41$  \\
         \end{tabular}         
         &         
          \begin{tabular}{c|c|c|c}
          $n$ & 3 & 4 & 5 \\ 
          \hline 
         $\left\langle E_{12}(n,p)\right\rangle$ & $0.33$  & $0.40$  & $0.42$  \\
          \hline
         $\sigma_n(p)$ &   $0.12$  & $0.08$ & $0.05$  \\
         \hline
        $\eta_n(p)$ & $-0.19$  &$-0.42$  & $-0.37$  \\
         \end{tabular}             
         &         
          \begin{tabular}{c|c|c|c}
         $n$ &  3 & 4 & 5 \\ 
          \hline 
         $\left\langle E_{12}(n,p)\right\rangle$ &  $0.23$  &$0.28$  &  $0.30$ \\
          \hline
        $\sigma_n(p)$ &   $0.09$  & $0.06$ & $0.04$  \\
         \hline
        $\eta_n(p)$ &   $-0.07$  & $-0.30$ & $-0.27$  \\
         \end{tabular}       \\  
%         &         
%          \begin{tabular}{c|c|c}
%          3 & 4 & 5 \\ 
%          \hline 
%          $0.14$  &$0.18$  &  $0.19$ \\
%          \hline
%         $0.07$  &$0.05$  &$0.03$   \\
%         \hline
%         $0.18$  & $-0.05$ & $-0.11$  \\
%         \end{tabular}        
%         &        
%          \begin{tabular}{c|c|c}
%          3 & 4 & 5 \\ 
%          \hline 
%          $0.07$  &$0.10$  & $0.11$  \\
%          \hline
%         $0.05$ & $0.04$ & $0.02$   \\
%         \hline
%         $0.55$ &$0.28$  & $0.16$  \\
%         \end{tabular}  \\
\hline                                             
\end{tabular}\\

\end{tabular}
\caption{\textbf{Statistical quantities  of the normalized frequency distributions under local and global noise.} Tabulation of mean, standard deviation, and skewness of the normalized frequency distributions corresponding to the three-, four-, and five-qubit random pure states subjected to the depolarizing, phase-flip, amplitude-damping, and white noise for noise strengths given by $p=0.1,0.2,0.3$. The sample size considered for the determination of the metrics is $N_S=5\times 10^4$ for each of the cases. All quantities are dimensionless.}
\label{tab:mean_noise}
\end{table*}
\normalsize

We find that the NFD of $E_{12}^0(n)$ in the case of states belonging to three-qubit $W$ class is qualitatively different from the  GHZ class, as is evident from  Fig.~\ref{fig:nonoise}. In contrast to the high value $(>0.5)$ of the mean of the NFD in the case of the GHZ class, $\langle E_{12}^0(n)\rangle$ has a comparatively lower value for the states from the W class.  It is also clear from Fig.~\ref{fig:nonoise} that the shapes of the distributions obtained from these two classes are strikingly different which can be confirmed from the  values of the SD and the skewness of the respective NFDs (see Table~\ref{tab:mean_nonoise}). It also indicates that the distributions derived from the states belonging to the set of  measure zero can show certain trait that cannot be seen for random pure states. This result is similar to the difference between the GHZ and the W class states in terms of the monogamy scores~\cite{giorgi2011,*prabhu2012,*dhar2017} of quantum correlation measures, which are also considered to be multiparty in nature. This also strengthens the potential of the LE computed over a pair of qubits to be considered as a multiparty measure of entanglement.

The above discussion indicates that it can also be interesting to investigate the distribution of LE for a certain family of states with higher number of qubits.  For high value of $n$, several such family  of states  exists. For our investigation, we randomly generate four-qubit genneralized Dicke states~\cite{dicke1954,*kumar2017,*bergmann2013,*lucke2014,*chiuri2012} with a single and double excitations, given by
\begin{eqnarray}
\ket{\psi_n^r} =  \sum_{i} a_i\mathcal{P}_i\left[ \ket{0}^{\otimes n-r}\ket{1}^{\otimes r}\right],
\label{eq:dicke}
\end{eqnarray}
with $a_i=\alpha_i+\text{i}\beta_i$ being complex numbers ($\alpha_i,\beta_i$ are real numbers chosen from  normal distributions of vanishing mean and unit SD -- a procedure similar to the case of the random $n$-qubit pure states and the three-qubit  W class states) such that $\sum_i|a_i|^2=1$, and  the summation in Eq.~(\ref{eq:dicke}) being over all possible permutations, $\{\mathcal{P}_i\}$, of the product state $ \ket{0}^{\otimes n-r}\ket{1}^{\otimes r}$ having $r$ qubits in the excited state, $\ket{1}$, and the rest of the qubits  in the ground state, $\ket{0}$. Like the W class states, these Dicke states also form a set of measure zero in the state-space of four-qubit systems for a reason similar to that in the case of the three-qubit W class states. Moreover, the Haar uniformity of these randomly generated Dicke states can also be proven via a procedure similar to that in the case of W states.   We observe that the NFD of $E_{12}^0(n)$ corresponding to random pure states of the form (\ref{eq:dicke}) with $n=4$, $r=1$ is similar to that of the W class states, as in Fig.~\ref{fig:nonoise}, while the same corresponding to $n=4$, $r=2$ has a different shape which is almost identical to the NFD corresponding to the three-qubit GHZ class states. With respect to the above observation, we also notice that in the case of the three-qubit GHZ class states and the four-qubit Dicke states with two excitations, post-measurement states of the two qubits on which entanglement is localized, correspondoing to the optimal measurement set-up, in both the cases have comparable non-zero values of entanglement. On the other hand, in the case of the three-qubit W class states, only one of the two-qubit post-measurement states corresponding to the optimal measurement has non-zero entanglement value, while the other post-measurement state has almost vanishing entanglement. This is similar to the case of the four-qubit Dicke states with one excitation, where corresponding to the optimal measurement setting, only one of the post-measurement states have non-zero entanglement while the entanglement values fo the other post-measurement two-qubit states vanish.

\subsection{Effects of noise}
\label{subsec:le_noise}

Upto now, distribution of LE has been considered for Haar uniformly chosen multi-qubit quantum states in a noiseless scenario.  However, in a realistic situation, a multi-party state, shared between several parties at different locations, can almost always be affected by  noise. It is, therefore,  of practical interest to check how the above results are modified in presence of different types of noise. In order to observe the consequence of noise on the distribution of LE,  we first generate random pure states $\rho_0=\ket{\psi}\bra{\psi}$ of three-, four- and five-qubits. Then, all the qubits are affected by a specific noisy channel, as described in Sec.~\ref{subsec:noise}. Specifically, for a   fixed value of the noise parameter $p$, we determine the NFD of $E_{12}(n,p)$ given by 
\begin{eqnarray}
f_n^p=\frac{N({E_{12}(n,p)})}{N_S},
\end{eqnarray}
where we compute $E_{12}(n,p)$ as mentioned earlier (see discussion succeeding Eq.~(\ref{eq:nfd_0})).
%This is determined by computing the values of $E_{12}(p,N)$ for the random  $n$-qubit states $\rho_n(p)$ obtained by applying noisy channels to a sample of $N_S$ random pure states, and counting the frequency $N({E_{12}(n,p)})$ of $E_{12}(n,p)$. 
Similar to the noiseless scenario, in the case of three-qubit systems, we separately generate the pure states belonging to the W class, and investigate the effect of noise on LE in these states, as presented in the subsequent discussion.

\small 
\begin{table*}[ht]
\begin{tabular}{cc}
\begin{tabular}{c}
Depolarizing noise \\
\begin{tabular}{|c|c|c|c|c|c|}
\hline
NFD Metric & $p=0.1$ &  $p=0.2$ & $p=0.3$ & $p=0.4$ & $p=0.5$ \\
\hline 
$\left\langle E_{12}(n,p)\right\rangle$ & $0.15$ & $0.05$ & $0.01$ & $0.00$ & $0.00$  \\
\hline 
$\sigma_n(p)$ & $0.13$ & $0.06$ &  $0.01$ & $0.00$ & $0.00$  \\
\hline 
$\eta_n(p)$ & $0.82$ & $1.27$ & $2.44$ & $15.45$ & $0.00$ \\
\hline 
\end{tabular} \\
\end{tabular}  &
\begin{tabular}{c}
Phase-flip noise \\
\begin{tabular}{|c|c|c|c|c|c|}
\hline
NFD Metric & $p=0.1$ &  $p=0.2$ & $p=0.3$ & $p=0.4$ & $p=0.5$ \\
\hline 
$\left\langle E_{12}(n,p)\right\rangle$ & $0.23$ &  $0.16$ & $0.10$ & $0.06$ & $0.03$  \\
\hline 
$\sigma_n(p)$ & $0.16$ & $0.11$ &  $0.07$ & $0.04$ & $0.02$  \\
\hline 
$\eta_n(p)$ & $0.62$ & $0.64$ & $0.67$ & $0.68$ & $0.72$ \\
\hline 
\end{tabular} \\
\end{tabular}\\
\begin{tabular}{c}
Amplitude-damping noise \\
\begin{tabular}{|c|c|c|c|c|c|}
\hline
NFD Metric & $p=0.1$ &  $p=0.2$ & $p=0.3$ & $p=0.4$ & $p=0.5$ \\
\hline 
$\left\langle E_{12}(n,p)\right\rangle$ & $0.26$ &  $0.21$ & $0.17$ & $0.13$  & $0.10$  \\
\hline 
$\sigma_n(p)$ & $0.18$ & $0.15$ &  $0.13$ & $0.10$ & $0.07$  \\
\hline 
$\eta_n(p)$ & $0.63$ &  $0.67$ & $0.72$ & $0.74$ & $0.79$ \\
\hline 
\end{tabular} \\
\end{tabular} &
\begin{tabular}{c}
White Noise \\
\begin{tabular}{|c|c|c|c|c|c|}
\hline
NFD Metric & $p=0.1$ &  $p=0.2$ & $p=0.3$ & $p=0.4$ & $p=0.5$ \\
\hline 
$\left\langle E_{12}(n,p)\right\rangle$ &  $0.25$ &  $0.18$ & $0.13$ & $0.08$  & $0.04$  \\
\hline 
$\sigma_n(p)$ & $0.18$ & $0.15$ &  $0.12$ & $0.08$ & $0.05$  \\
\hline 
$\eta_n(p)$ & $0.66$ &  $0.76$ & $0.91$ & $1.10$ & $1.45$ \\
\hline 
\end{tabular} \\
\end{tabular} \\
\end{tabular} 
\caption{\textbf{Metrics of the normalized frequency distributions for states from three-qubit W class under noise.} Tabulation of the mean, standard deviation, and skewness of the normalized frequency distributions corresponding to the three-qubit random pure states belonging to the W class, subjected to the depolarizing, phase-flip, amplitude-damping, and white noise for five specific noise strengths given by $p=0.0,0.2,0.3,0.4,$ and $0.5$. The sample size is same as Table~\ref{tab:mean_noise}. All quantities are dimensionless.}
\label{tab:W_mean_noise}
\end{table*} 
\normalsize

We first concentrate on the three-qubit GHZ class along with the Haar uniformly generated random pure states of four- and five-qubits, subjected to local noise. For $p>0$, we evaluate $f_n^p$ for fixed values of $p$, with $n=3,4,5$. The profiles of $f_n^p$ corresponding to $n=3,4$, for five different noise strengths $p=0.1,0.2,0.3,0.4,0.5$, and for the different types of noise (see Sec.~\ref{subsec:noise}) are depicted in Fig.~\ref{fig:noise}, while the metrics of the NFDs are tabulated in Table~\ref{tab:mean_noise}.  The profile of $f_n^p$ corresponding to $n=5$ are similar to that of $n=4$, and the GHZ class states for $n=3$.  Since the local noise considered in this paper are Markovian, one expects $E_{12}(n,p_2)\leq E_{12}(n,p_1)\leq E_{12}^0(n)$, where $1\geq p_2\geq p_1\geq 0$, for a fixed value of $n$. In agreement with this,  for a fixed $n$, the mean, $\left\langle E_{12}(n,p)\right\rangle$, of the NFDs satisfy 
\begin{eqnarray}
\left\langle E_{12}(n,p_2)\right\rangle\leq \left\langle E_{12}(n,p_1)\right\rangle \leq \left\langle E_{12}^0(n)\right\rangle
\label{eq:nfd_local_p}
\end{eqnarray}
for $1\geq p_2\geq p_1\geq 0$, and the peak of the distribution shifts towards lower values of $E_{12}(n,p)$ with increasing $p$, as is evident from Fig.~\ref{fig:noise}. On the other hand, for a fixed $p$, we interestingly find
\begin{eqnarray}
\left\langle E_{12}(n_1,p)\right\rangle\approx\left\langle E_{12}(n_2,p)\right\rangle
\label{eq:nfd_local_n}
\end{eqnarray}
for $n_1>n_2$, $n_1,n_2\in\{3,4,5\}$, for a specific type of noise, where the difference between $\left\langle E_{12}(n_1,p)\right\rangle$ and $\left\langle E_{12}(n_2,p)\right\rangle$ occurring only in the second decimal place. This is in  sharp contrast to the finding for the noiseless scenario, as summarized in Eq.~(\ref{eq:compare_1}).   The values of the SD, $\sigma_n^p$ (with a definition similar to that in Eq.~(\ref{eq:sd}) with a non-zero $p$), of the NFDs are found to  satisfy 
\begin{eqnarray}
\sigma_{n}(p_1)&\geq &\sigma_{n}(p_2)\;\text{for}\; 1\geq p_2>p_1\geq 0,\nonumber \\ 
\sigma_{n_1}(p)&\geq &\sigma_{n_2}(p)\;\text{for}\; n_2>n_1;\;n_1,n_2\in\{3,4,5\},
\label{eq:nfd_local_sd}
\end{eqnarray} 
for a fixed type of noise. 

We now move on to the comparison of LE between different noise models. While DP noise  turns out to be the most stringent in pushing the average value of LE towards zero, interestingly, considerably high values of localizable entanglement sustain for the AD noise when the same noise strength $p$ as in the case of the DP and the PF noise is applied. This is clearly visible from the plots in Fig.~\ref{fig:noise} and  Table~\ref{tab:mean_noise}. Also, the values of the metrics as well as the profiles of the NFDs suggest that the effect of noise on the distribution of LE is qualitatively similar for the PF and the AD channels, while being considerably different from the same corresponding to the DP channel.  In order to obtain a full picture about the complete state-space of the three-qubit system, we separately generate three-qubit states belonging to the W class, and perform the same analysis as in the case of states belonging to the three-qubit GHZ class (see Fig.~\ref{fig:noise} and Table~\ref{tab:mean_noise}). 
%The profiles of the corresponding NFDs are exhibited in Fig.~\ref{fig:noise}, while the values of the metrics of the NFDs are tabulated in Table~\ref{tab:W_mean_noise}. 
The profiles of the NFDs clearly indicate a higher robustness of the multi-qubit quantum states of three- (including the GHZ and the W class states), four-, and five-qubits towards AD noise as far as the value of LE is concerned, in comparison to the PF and the DP noise. We will present a more quantitative analysis on this topic in Sec.~\ref{sec:robust}.  In order to check how the NFDs corresponding to the zero-measure states for $n>3$ response against local noise, we apply PF, DP, and AD noise on the qubits of four-qubit generalized  Dicke states with single and double excitations (Eq.~(\ref{eq:dicke})),  and determine the NFDs of $E_{12}(n,p)$ for different values of $p>0$ for each of the noise-types. We find that our observation regarding the difference between the shapes of the NFDs corresponding to the Dicke states with single and double excitations in noiseless case holds in the noisy scenario also.

\begin{figure}
\includegraphics[scale=0.75]{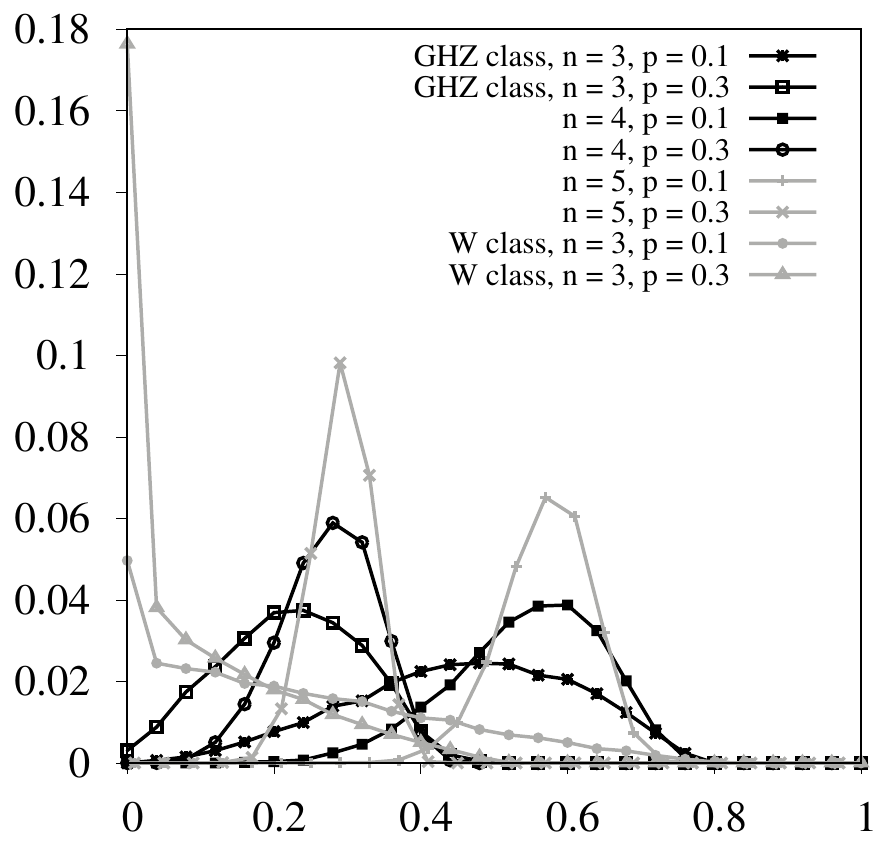}
\caption{(Color online.) \textbf{Normalized frequency distribution under white noise.}  $f_n^p$ (vertical axes) with respect to $E_{12}(n,p)$ (horizontal axes). Random pure states in the case of $n=3,4,5$, and the three-qubit W class states are affected by white noise of strengths $p=0.1,0.3$. The sample size is the same as in Fig.~\ref{fig:noise}. All quantities plotted are dimensionless.}
\label{fig:whnoise}
\end{figure}

To see the effect of  global noise, we investigate the  
%on the NFDs, we perform the analysis considering white noise on all the qubits in the $n$-qubit system, as given in Eq.~(\ref{eq:whitenoise}). The corresponding 
profiles of NFDs for a specific noise strengths as depicted in Fig.~\ref{fig:whnoise}). Comparing Tables~\ref{tab:mean_noise} and \ref{tab:W_mean_noise}, the states in the GHZ class are more robust 
% It is clear from Fig.~\ref{fig:whnoise} that in the case of three-qubit systems, GHZ class is more robust 
 to the white noise than that from the W class.  
% Also, the NFDs corresponding to the $4$- and $5$-qubit systems are qualitatively similar with the same for the $3$-qubit GHZ class states under white noise, as is evident from the figure. The metrics of the different NFDs corresponding to the white noise are tabulated in Table~\ref{tab:mean_noise}. The values for the three-qubit GHZ class states and the random $4$- and $5$-qubit states suggest  trends of  $\left\langle E_{12}(n,p)\right\rangle$ and $\sigma_n(p)$ as functions of $n$ (for fixed $p$) and $p$ (for fixed $n$) similar to those corresponding to the local noise, as shown in Eqs.~(\ref{eq:nfd_local_p})-(\ref{eq:nfd_local_sd}).   
Similar to the local noise, here also we observe that for fixed noise strength, mean and SD of NFDs do not change with the increase of $n$ (see Table~\ref{tab:mean_noise}).

\begin{figure*}
\includegraphics[width=\textwidth]{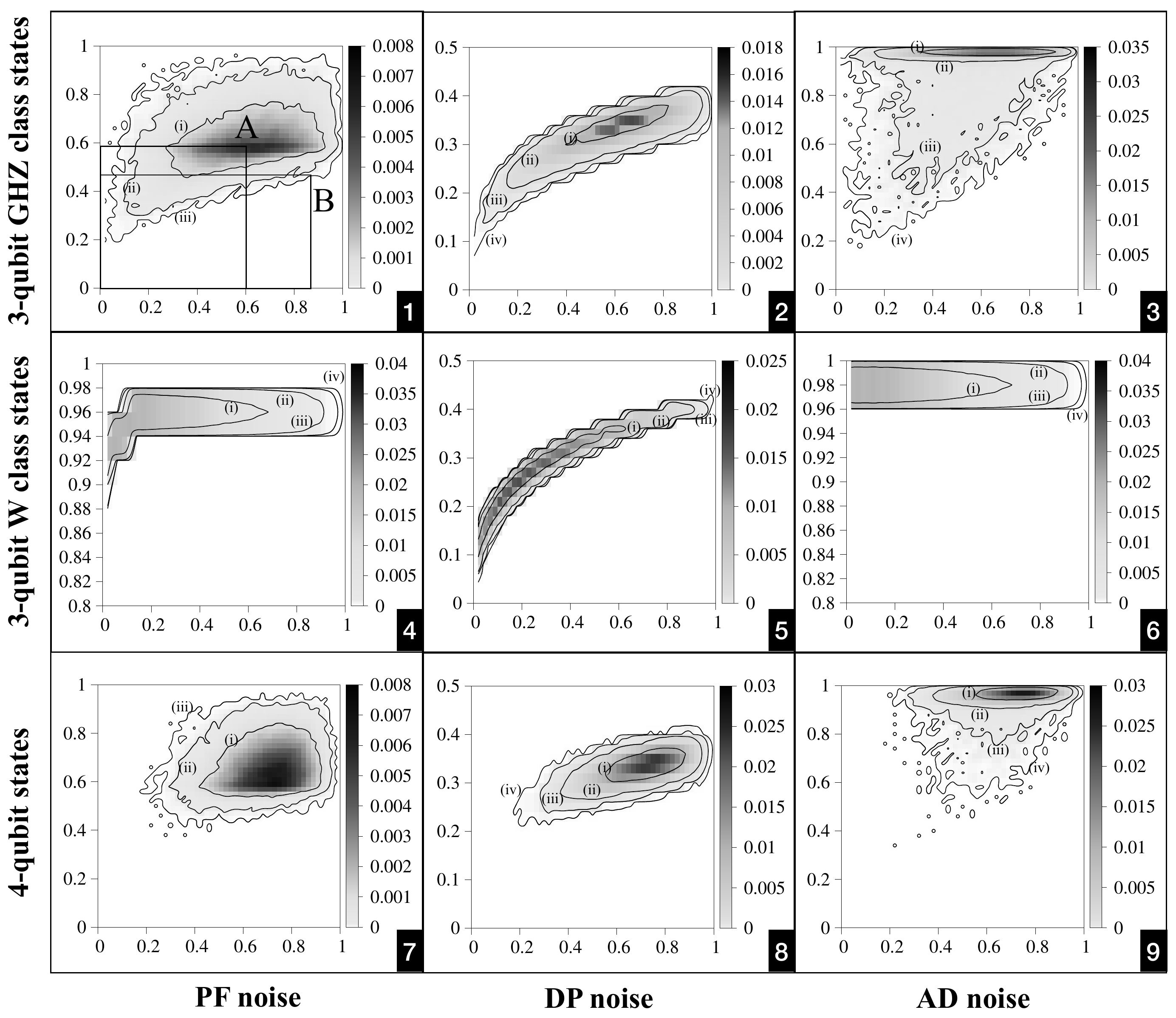}
\caption{(Color online.) Contour map of $E_L^0$ and $p_c$ under local noise. The rows and columns of the figure and the sample size used to generate the data are the same as in Fig.~\ref{fig:noise}. The values of $f\left(E_{12}^0(n),p_c\right)$ corresponding to the different contour lines are as follows. \textbf{[1],[7]:} $10^{-3}$ (i), $10^{-4}$ (ii), $10^{-5}$ (iii), from inside to outside, \textbf{[2-6,8,9]:} $10^{-2}$ (i), $10^{-3}$ (ii), $10^{-4}$ (iii), $10^{-5}$ (iv), from inside to outside. All quantities plotted are dimensionless.} 
%Normalized frequency distribution $f\left(E_L^{0},p_c\right)$ (axes perpendicular to the page) against $E_L^{0}$ (horizontal axes on the plane of the page) and $p_c$ (vertical axes on the plane of the page) for Haar uniformly generated random pure states in the case of $N=3$ (GHZ and W class) and $4$ (from top to bottom horizontal panels), subjected to phase-flip (left column), depolarizing (middle column), and amplitude-damping (right column) noise.   A sample of $N_S=5\times 10^{4}$ random pure initial ($p=0$) states are considered for each of the normalized frequency distributions. All quantities plotted are dimensionless.}
\label{fig:dpcr}
\end{figure*}

\section{Robustness against noise}
\label{sec:robust}

We have pointed out that the effect of noise on the distribution of localizable entanglement is qualitatively similar for the PF and the AD channels, although  LE seems to survive against more noise  in the case of the AD channel. In this section, we aim for an unambiguous conclusion on the robustness of a multi-qubit quantum state, in terms of its LE, against different types of noise.  In order to formulate a figure of merit for this purpose, we look at two specific quantities -- (1) the initial value of localizable entanglement, $E_{12}^0(n)$, of the $n$-qubit state,  and (2) the value of  noise strength, $p_c$, which we refer to as the \emph{critical strength of noise}, such that 
\begin{eqnarray}
 E_{12}(n,p)&>& 0,\text{for }p<p_c,\nonumber \\
 E_{12}(n,p)&=& 0,\text{for }p\geq p_c,
 \label{eq:def_pc}
\end{eqnarray}
provided $E_{12}^0(n)>0$. Whether a high (low) value of $E_{12}^0(n)$ implies a high (low) value of $p_c$, or whether $p_c$ has any other functional dependence  on the value of $E_{12}^0(n)$, are non-trivial questions, and need careful consideration for a quantitative analysis of the robustness of localizable entanglement against noise. 

Towards this goal, we Haar uniformly simulate a sample of random pure states of size $N_S$ for each of the three-, four-, five-qubit systems, and apply different types of local as well as global noise on the qubits. We define the normalized frequency 
\begin{eqnarray}
f\left(E_{12}^0(n),p_c\right)=\frac{N(E_{12}^0(n),p_c)}{N_S},
\end{eqnarray} 
where $N(E_{12}^0(n),p_c)$ is the number of Haar uniformly generated random pure states with $E_L^0(n)$ satisfying Eq.~(\ref{eq:def_pc}) for a system of $n$ qubits. For $N_S=5\times 10^4$, we plot $f\left(E_{12}^0(n),p_c\right)$ as functions of $E_{12}^0(n)$ and $p_c$ in Fig.~\ref{fig:dpcr} for Haar uniformly generated  three- and four-qubit random pure states subjected to local noise, where the bin size for determining $f\left(E_{12}^0(n),p_c\right)$ is taken to be  $10^{-1}\times 10^{-1}$.  The observations emerging from Fig.~\ref{fig:dpcr} are as follows.

%\begin{figure*}
%\includegraphics[width=0.8\textwidth]{whdpcr.pdf}
%\caption{(Color online.) \textbf{Normalized frequency distribution of $E_L^0$ and $p_c$ for white noise.} Normalized frequency distribution $f\left(E_L^{0},p_c\right)$ (axes perpendicular to the plane of the page) against $E_L^{0}$ (horizontal axes on the plane of the page) and $p_c$ (vertical axes on the plane of the page) for Haar uniformly generated random pure states in the case of $N=3,4,5$, subjected to white noise. A sample of $N_S=5\times 10^{4}$ random pure initial ($p=0$) states are considered for each normalized frequency distribution. All quantities plotted are dimensionless.}
%\label{fig:whdpcr}
%\end{figure*}

%\begin{figure*}
%\includegraphics[width=0.9\textwidth]{gen.pdf}
%\caption{(Color online.) \textbf{Scatter-plot of  $p_c$ against $E_L^0$ for three-qubit states under local noise.} Scatter-plots of  $p_c$ (vertical axes) as a function of $E_L^{0}$ (horizontal axes)  for Haar uniformly generated random three-qubit pure states belonging to GHZ and W class, and state having the form of genenralized GHZ and W states  subjected to phase-flip (left column), depolarizing (middle column), and amplitude-damping (right column) noise.   A sample of $N_S=5\times 10^{4}$ random pure initial ($p=0$) states are considered for each of the scatter plots. All quantities plotted are dimensionless.}
%\label{fig:scatter}
%\end{figure*}

\noindent\textbf{(1)} In the case of random pure multi-qubit $(n=3,4,5)$ states with  a  fixed initial value of $E_{12}^0(n)$, there can be a number of values of $p_c$, as indicated from the vertical spreads of the $f\left(E_{12}^0(n),p_c\right)$ in  Fig.~\ref{fig:dpcr}. It clearly indicates that the critical value of noise-strength is independent of the content of the LE of the initial state. Also, a relatively low value of $E_{12}^0(n)$ may correspond to a higher value of $p_c$ compared to the same for a high value of $E_{12}^0(n)$ -- see, for example,  the points A and B in the case of three-qubit GHZ class states under PF noise. 

\noindent\textbf{(2)} There is a qualitative difference between the landscape of $f\left(E_{12}^0(n),p_c\right)$ in the case of three-qubit GHZ and W class states under PF noise. While the former indicates existence of the values of $E_{12}^0(n)$ and $p_c$ over a broad range in the region $0\leq E_{12}^0(n)\leq 1$, $0\leq p_c\leq 1$, in the latter, most of the values of $p_c$ are confined in the range $0.9\leq p_c\leq 1$, while the values of $E_{12}^0(n)$ is consistent with the profiles of the NFD corresponding to the three-qubit W class states in the noiseless situation (see Fig.~\ref{fig:nonoise}). 

\noindent\textbf{(3)} The NFDs, $f\left(E_{12}^0(n),p_c\right)$, corresponding to the three- (GHZ class), four-, and five-qubit random pure states under local noise exhibit qualitatively similar trend. The concentration of the randomly chosen states shifts towards higher values of $E_{12}^0(n)$ when  $n$ increases, which is consistent with the findings in Sec.~\ref{subsec:le_nonoise} (see Eq.~(\ref{eq:compare_1}) and Fig.~\ref{fig:nonoise}).

\noindent\textbf{(4)} As predicted in Sec.~\ref{subsec:le_noise}, the DP noise turns out to be the most destructive one, which is evident from the considerably low values of $p_c$ irrespective of the values of $E_{12}^0(n)$, among all types of local noise as well as white noise  and random initial states. 

\noindent\textbf{(5)} In the case of the AD noise, irrespective of the values of $E_{12}^0(n)$, the values of $p_c$ for majority of the three-qubit W class random pure states are found to be in the range $0.9\leq p_c\leq 1$ -- a feature similar to the W class states under PF noise. On the other hand, for the  three-qubit GHZ class states,  the four- and the five-qubit random pure states under AD noise, there is a large fraction of states for which $ p_c \in [0.9,1]$, although there are  fraction of states scattered over  $0\leq p_c < 0.9$.  This indicates (1) a higher robustness of LE of multi-qubit random three-qubit pure states belonging to the W class  under the influence of AD and PF noise, and (2) a higher robustness of LE of a large fraction of three-qubit GHZ class states as well as four- and five-qubit random pure states under AD noise. These findings are in a good agreement with the inference in Sec.~\ref{subsec:le_noise}.   

The trends of the NFD, $f\left(E_{12}^0(n),p_c\right)$, corresponding to the three-, four-, and five-qubit random pure states under white noise is similar to the DP noise, which is  shown in Fig.~\ref{fig:dpcr}. In all other cases, the change in the landscape is as per the change in the distribution of $E_{12}^0(n)$ when $n$ increases from $3$ to $5$ (see Fig.~\ref{fig:whnoise}). 

Let us now check the status of $p_c$ of two important families of three-qubit states, namely, 
%It is interesting to  to check whether the zero-measure states in the state space of three qubits, such as 
the generalized GHZ~\cite{greenberger1989} and generalized W~\cite{zeilinger1992, dur2000}  states. The former is defined as 
\begin{eqnarray}
\ket{\psi}=a_0\ket{000}+a_1\ket{111},
\end{eqnarray}
with $|a_0|^2+|a_1|^2=1$, $a_0$ and $a_1$ being complex numbers having the form $a_j=\alpha_j+\text{i}\beta_j$, $j=0,1$, where $\alpha_j$ and $\beta_j$ are real numbers. On the other hand, a three-qubit  generalized W state $\ket{\psi}$ has the form 
\begin{eqnarray}
\ket{\psi}=a_0\ket{001}+a_1\ket{010}+a_2\ket{100}, 
\end{eqnarray}
with $\sum_{j=0}^{2}|a_j|^2=1$. The state parameters $a_j$, $i=0,1,2$, are complex numbers having the form $a_j=\alpha_j+\text{i}\beta_j$, $j=0,1,2$, where $\alpha_j$ and $\beta_j$ are real numbers. Specifically, here we see whether the generalized GHZ and W  states exhibit behaviours different than those shown by the GHZ class and the W class states respectively,  under the application of noise. Towards this aim,  we investigate the variation of the values of $p_c$ against that of  $E_{12}^0(n)$ for the 
%Haar uniformly simulated random three-qubit states in the form of  
generalized GHZ and the W states subjected to local noise channels.   We find that the trends of $p_c$ as a function of $E_{12}^0(n)$ are similar, qualitatively as well as quantitatively,  for the generalized W states and the states belonging to the W class for all the three types of noise.  
However, such similarities are not always present between the generalized GHZ states and the states from the GHZ class. 
%In the case of three-qubit generalized GHZ states under local noise, the trends of $p_c$ against $E_{12}^0(n)$ corresponding to the PF noise differs from the same corresponding to the DP noise, when considered together with the results obtained from the GHZ class states. 
Let us denote the critical noise strength $p_c$ corresponding to a generalized GHZ state having initial LE $E_{12}^0(n)$ by $p_c^{(1)}$, and the same corresponding to a GHZ class state having the same initial LE $E_{12}^0(n)$ by $p_c^{(2)}$. In the case of the PF channel and for all values of $E_{12}^0(n)$, for majority of the GHZ class states, $p_c^{(2)}\leq p_c^{(1)}$, while in the case of the DP noise, there exists considerable number of states in GHZ class for which $p_c^{(2)}\geq p_c^{(1)}$ for all  $E_{12}^0(n)$. For DP noise, the fraction of states in GHZ class for which $p_c^{(2)}\geq p_c^{(1)}$ increases with increasing $E_{12}^0(n)$. The results again confirms that the properties of  random pure states cannot be mimicked by a family or subset of states.

\vspace{0.2cm}

\section{Conclusion}
\label{sec:conclusion}

To summarize, we investigated whether multi-qubit random pure states can exhibit high values of  localizable entanglement (LE) concentrated over a chosen pair of qubits. Due to the computational limitation imposed by the difficulty in achieving the maximization involved in the definition of localizable entanglement, we have restricted our study in systems composed of three-, four-, and five-qubits. By determining the normalized frequency distributions, we showed that for Haar uniformly generated random pure states, the average value of localizable entanglement  increases with increasing the number of qubits in the system. Also, high clustering of the randomly generated multi-qubit quantum states around higher values of localizable entanglement is signalled by other metrics of the normalized frequency distribution, such as the standard deviation and the skewness.  This feature bears similarity with the characteristics of genuine multi-party entanglement as shown in previous works. It also indicates that  LE can mimic properties similar to a valid multi-party entanglement measure. 

In order to check how this characteristic changes in a realistic scenario when noise is introduced in the system, we apply phase-flip, depolarizing, amplitude damping, and white noises on the Haar uniformly simulated random pure states of three-, four-, and five-qubit systems, and study the variation of the metrics of the normalized frequency distribution with varying noise strength. We found that the mean of LE does not increase with the increase of the number of parties for a fixed noise strength. Instead, it remains almost constant for a fixed value of noise which is in contrast with the noiseless situation. Such a feature is independent of the choices of noise models considered in this paper. Investigation also reveals that amplitude damping channel destroys less LE compared to any other channels. Moreover, we find that the critical noise strength above which LE vanishes does not depend on the initial LE of a given state. For the amplitude and phase damping noise, the analysis of LE shows that the states from the W class is more robust than that from the GHZ class.     

%We point out that for a non-zero noise strength, the difference between the average values of localizable entanglement over a pair of qubits in multi-qubit systems with three, four, and five qubits diminishes, and even vanishes for certain values of noise strength. We also comment on the robustness of localizable entanglement against different types of noise considered in this paper. Our results indicate that localizable entanglement over a qubit-pair in random three-, four, and five-qubit pure  states has a relatively high robustness against the amplitude damping noise. 

Our analysis also reveals that under decoherence, nature of entanglement content of most of the multi-qubit states are similar and not maximal unlike the noiseless scenario. If such patterns in presence of noise  persists for other multi-party entanglement measures,  they may be useful for quantum computational speed-up, especially in a realistic scenario. Our study also highlights the potential of localizable entanglement to be considered as an appropriate candidate in revealing  the multi-party nature of quantum correlation present in a composite quantum system, and motivates extensive research in this direction from the perspective of quantum information processing tasks.

\acknowledgments
The authors acknowledge computations performed at the cluster computing facility of Harish-Chandra Research Institute, Allahabad, India. RB acknowledges the use of \href{https://github.com/titaschanda/QIClib}{QIClib} -- a modern C++11 library for general purpose quantum computing. The authors thank Ujjwal Sen for fruitful discussions.

\bibliography{lib3}

\begin{thebibliography}{98}%
\makeatletter
\providecommand \@ifxundefined [1]{%
 \@ifx{#1\undefined}
}%
\providecommand \@ifnum [1]{%
 \ifnum #1\expandafter \@firstoftwo
 \else \expandafter \@secondoftwo
 \fi
}%
\providecommand \@ifx [1]{%
 \ifx #1\expandafter \@firstoftwo
 \else \expandafter \@secondoftwo
 \fi
}%
\providecommand \natexlab [1]{#1}%
\providecommand \enquote  [1]{``#1''}%
\providecommand \bibnamefont  [1]{#1}%
\providecommand \bibfnamefont [1]{#1}%
\providecommand \citenamefont [1]{#1}%
\providecommand \href@noop [0]{\@secondoftwo}%
\providecommand \href [0]{\begingroup \@sanitize@url \@href}%
\providecommand \@href[1]{\@@startlink{#1}\@@href}%
\providecommand \@@href[1]{\endgroup#1\@@endlink}%
\providecommand \@sanitize@url [0]{\catcode `\\12\catcode `\$12\catcode
  `\&12\catcode `\#12\catcode `\^12\catcode `\_12\catcode `\%12\relax}%
\providecommand \@@startlink[1]{}%
\providecommand \@@endlink[0]{}%
\providecommand \url  [0]{\begingroup\@sanitize@url \@url }%
\providecommand \@url [1]{\endgroup\@href {#1}{\urlprefix }}%
\providecommand \urlprefix  [0]{URL }%
\providecommand \Eprint [0]{\href }%
\providecommand \doibase [0]{http://dx.doi.org/}%
\providecommand \selectlanguage [0]{\@gobble}%
\providecommand \bibinfo  [0]{\@secondoftwo}%
\providecommand \bibfield  [0]{\@secondoftwo}%
\providecommand \translation [1]{[#1]}%
\providecommand \BibitemOpen [0]{}%
\providecommand \bibitemStop [0]{}%
\providecommand \bibitemNoStop [0]{.\EOS\space}%
\providecommand \EOS [0]{\spacefactor3000\relax}%
\providecommand \BibitemShut  [1]{\csname bibitem#1\endcsname}%
\let\auto@bib@innerbib\@empty
%</preamble>
\bibitem [{\citenamefont {Horodecki}\ \emph {et~al.}(2009)\citenamefont
  {Horodecki}, \citenamefont {Horodecki}, \citenamefont {Horodecki},\ and\
  \citenamefont {Horodecki}}]{horodecki2009}%
  \BibitemOpen
  \bibfield  {author} {\bibinfo {author} {\bibfnamefont {R.}~\bibnamefont
  {Horodecki}}, \bibinfo {author} {\bibfnamefont {P.}~\bibnamefont
  {Horodecki}}, \bibinfo {author} {\bibfnamefont {M.}~\bibnamefont
  {Horodecki}}, \ and\ \bibinfo {author} {\bibfnamefont {K.}~\bibnamefont
  {Horodecki}},\ }\href {\doibase 10.1103/RevModPhys.81.865} {\bibfield
  {journal} {\bibinfo  {journal} {Rev. Mod. Phys.}\ }\textbf {\bibinfo {volume}
  {81}},\ \bibinfo {pages} {865} (\bibinfo {year} {2009})}\BibitemShut
  {NoStop}%
\bibitem [{\citenamefont {Bennett}\ \emph {et~al.}(1993)\citenamefont
  {Bennett}, \citenamefont {Brassard}, \citenamefont {Cr\'epeau}, \citenamefont
  {Jozsa}, \citenamefont {Peres},\ and\ \citenamefont
  {Wootters}}]{bennett1993}%
  \BibitemOpen
  \bibfield  {author} {\bibinfo {author} {\bibfnamefont {C.~H.}\ \bibnamefont
  {Bennett}}, \bibinfo {author} {\bibfnamefont {G.}~\bibnamefont {Brassard}},
  \bibinfo {author} {\bibfnamefont {C.}~\bibnamefont {Cr\'epeau}}, \bibinfo
  {author} {\bibfnamefont {R.}~\bibnamefont {Jozsa}}, \bibinfo {author}
  {\bibfnamefont {A.}~\bibnamefont {Peres}}, \ and\ \bibinfo {author}
  {\bibfnamefont {W.~K.}\ \bibnamefont {Wootters}},\ }\href {\doibase
  10.1103/PhysRevLett.70.1895} {\bibfield  {journal} {\bibinfo  {journal}
  {Phys. Rev. Lett.}\ }\textbf {\bibinfo {volume} {70}},\ \bibinfo {pages}
  {1895} (\bibinfo {year} {1993})}\BibitemShut {NoStop}%
\bibitem [{\citenamefont {Bouwmeester}\ \emph {et~al.}(1997)\citenamefont
  {Bouwmeester}, \citenamefont {Pan}, \citenamefont {Mattle}, \citenamefont
  {Eibl}, \citenamefont {Weinfurter},\ and\ \citenamefont
  {Zeilinger}}]{bouwmeester1997}%
  \BibitemOpen
  \bibfield  {author} {\bibinfo {author} {\bibfnamefont {D.}~\bibnamefont
  {Bouwmeester}}, \bibinfo {author} {\bibfnamefont {J.-W.}\ \bibnamefont
  {Pan}}, \bibinfo {author} {\bibfnamefont {K.}~\bibnamefont {Mattle}},
  \bibinfo {author} {\bibfnamefont {M.}~\bibnamefont {Eibl}}, \bibinfo {author}
  {\bibfnamefont {H.}~\bibnamefont {Weinfurter}}, \ and\ \bibinfo {author}
  {\bibfnamefont {A.}~\bibnamefont {Zeilinger}},\ }\href {\doibase
  10.1038/37539} {\bibfield  {journal} {\bibinfo  {journal} {Nature}\ }\textbf
  {\bibinfo {volume} {390}},\ \bibinfo {pages} {575} (\bibinfo {year}
  {1997})}\BibitemShut {NoStop}%
\bibitem [{\citenamefont {Murao}\ \emph {et~al.}(1999)\citenamefont {Murao},
  \citenamefont {Jonathan}, \citenamefont {Plenio},\ and\ \citenamefont
  {Vedral}}]{murao1999}%
  \BibitemOpen
  \bibfield  {author} {\bibinfo {author} {\bibfnamefont {M.}~\bibnamefont
  {Murao}}, \bibinfo {author} {\bibfnamefont {D.}~\bibnamefont {Jonathan}},
  \bibinfo {author} {\bibfnamefont {M.~B.}\ \bibnamefont {Plenio}}, \ and\
  \bibinfo {author} {\bibfnamefont {V.}~\bibnamefont {Vedral}},\ }\href
  {\doibase 10.1103/PhysRevA.59.156} {\bibfield  {journal} {\bibinfo  {journal}
  {Phys. Rev. A}\ }\textbf {\bibinfo {volume} {59}},\ \bibinfo {pages} {156}
  (\bibinfo {year} {1999})}\BibitemShut {NoStop}%
\bibitem [{\citenamefont {Grudka}(2004)}]{grudka2004}%
  \BibitemOpen
  \bibfield  {author} {\bibinfo {author} {\bibfnamefont {A.}~\bibnamefont
  {Grudka}},\ }\href {https://arxiv.org/abs/quant-ph/0303112} {\bibfield
  {journal} {\bibinfo  {journal} {Acta Phys. Slov.}\ }\textbf {\bibinfo
  {volume} {54}},\ \bibinfo {pages} {291} (\bibinfo {year} {2004})},\ \Eprint
  {http://arxiv.org/abs/arXiv:quant-ph/0303112} {arXiv:quant-ph/0303112}
  \BibitemShut {NoStop}%
\bibitem [{\citenamefont {Sen(De)}\ and\ \citenamefont
  {Sen}(2010)}]{sende2010a}%
  \BibitemOpen
  \bibfield  {author} {\bibinfo {author} {\bibfnamefont {A.}~\bibnamefont
  {Sen(De)}}\ and\ \bibinfo {author} {\bibfnamefont {U.}~\bibnamefont {Sen}},\
  }\href {\doibase 10.1103/PhysRevA.81.012308} {\bibfield  {journal} {\bibinfo
  {journal} {Phys. Rev. A}\ }\textbf {\bibinfo {volume} {81}},\ \bibinfo
  {pages} {012308} (\bibinfo {year} {2010})}\BibitemShut {NoStop}%
\bibitem [{\citenamefont {Bennett}\ and\ \citenamefont
  {Wiesner}(1992)}]{bennett1992}%
  \BibitemOpen
  \bibfield  {author} {\bibinfo {author} {\bibfnamefont {C.~H.}\ \bibnamefont
  {Bennett}}\ and\ \bibinfo {author} {\bibfnamefont {S.~J.}\ \bibnamefont
  {Wiesner}},\ }\href {\doibase 10.1103/PhysRevLett.69.2881} {\bibfield
  {journal} {\bibinfo  {journal} {Phys. Rev. Lett.}\ }\textbf {\bibinfo
  {volume} {69}},\ \bibinfo {pages} {2881} (\bibinfo {year}
  {1992})}\BibitemShut {NoStop}%
\bibitem [{\citenamefont {Mattle}\ \emph {et~al.}(1996)\citenamefont {Mattle},
  \citenamefont {Weinfurter}, \citenamefont {Kwiat},\ and\ \citenamefont
  {Zeilinger}}]{mattle1996}%
  \BibitemOpen
  \bibfield  {author} {\bibinfo {author} {\bibfnamefont {K.}~\bibnamefont
  {Mattle}}, \bibinfo {author} {\bibfnamefont {H.}~\bibnamefont {Weinfurter}},
  \bibinfo {author} {\bibfnamefont {P.~G.}\ \bibnamefont {Kwiat}}, \ and\
  \bibinfo {author} {\bibfnamefont {A.}~\bibnamefont {Zeilinger}},\ }\href
  {\doibase 10.1103/PhysRevLett.76.4656} {\bibfield  {journal} {\bibinfo
  {journal} {Phys. Rev. Lett.}\ }\textbf {\bibinfo {volume} {76}},\ \bibinfo
  {pages} {4656} (\bibinfo {year} {1996})}\BibitemShut {NoStop}%
\bibitem [{\citenamefont {De}\ and\ \citenamefont {Sen}(2011)}]{sende2010}%
  \BibitemOpen
  \bibfield  {author} {\bibinfo {author} {\bibfnamefont {A.~S.}\ \bibnamefont
  {De}}\ and\ \bibinfo {author} {\bibfnamefont {U.}~\bibnamefont {Sen}},\
  }\href {https://arxiv.org/abs/1105.2412} {\bibfield  {journal} {\bibinfo
  {journal} {Phys. News}\ }\textbf {\bibinfo {volume} {40}},\ \bibinfo {pages}
  {17} (\bibinfo {year} {2011})},\ \Eprint
  {http://arxiv.org/abs/arXiv:1105.2412} {arXiv:1105.2412} \BibitemShut
  {NoStop}%
\bibitem [{\citenamefont {Bru\ss{}}\ \emph {et~al.}(2004)\citenamefont
  {Bru\ss{}}, \citenamefont {D'Ariano}, \citenamefont {Lewenstein},
  \citenamefont {Macchiavello}, \citenamefont {Sen(De)},\ and\ \citenamefont
  {Sen}}]{bruss2004}%
  \BibitemOpen
  \bibfield  {author} {\bibinfo {author} {\bibfnamefont {D.}~\bibnamefont
  {Bru\ss{}}}, \bibinfo {author} {\bibfnamefont {G.~M.}\ \bibnamefont
  {D'Ariano}}, \bibinfo {author} {\bibfnamefont {M.}~\bibnamefont
  {Lewenstein}}, \bibinfo {author} {\bibfnamefont {C.}~\bibnamefont
  {Macchiavello}}, \bibinfo {author} {\bibfnamefont {A.}~\bibnamefont
  {Sen(De)}}, \ and\ \bibinfo {author} {\bibfnamefont {U.}~\bibnamefont
  {Sen}},\ }\href {\doibase 10.1103/PhysRevLett.93.210501} {\bibfield
  {journal} {\bibinfo  {journal} {Phys. Rev. Lett.}\ }\textbf {\bibinfo
  {volume} {93}},\ \bibinfo {pages} {210501} (\bibinfo {year}
  {2004})}\BibitemShut {NoStop}%
\bibitem [{\citenamefont {Bru\ss{}}\ \emph {et~al.}(2006)\citenamefont
  {Bru\ss{}}, \citenamefont {Lewenstein}, \citenamefont {Sen(De)},
  \citenamefont {Sen}, \citenamefont {D'Ariano},\ and\ \citenamefont
  {Macchiavello}}]{bruss2006}%
  \BibitemOpen
  \bibfield  {author} {\bibinfo {author} {\bibfnamefont {D.}~\bibnamefont
  {Bru\ss{}}}, \bibinfo {author} {\bibfnamefont {M.}~\bibnamefont
  {Lewenstein}}, \bibinfo {author} {\bibfnamefont {A.}~\bibnamefont {Sen(De)}},
  \bibinfo {author} {\bibfnamefont {U.}~\bibnamefont {Sen}}, \bibinfo {author}
  {\bibfnamefont {G.~M.}\ \bibnamefont {D'Ariano}}, \ and\ \bibinfo {author}
  {\bibfnamefont {C.}~\bibnamefont {Macchiavello}},\ }\href {\doibase
  10.1142/S0219749906001888} {\bibfield  {journal} {\bibinfo  {journal} {Int.
  J. Quant. Info.}\ }\textbf {\bibinfo {volume} {4}} (\bibinfo {year} {2006}),\
  10.1142/S0219749906001888}\BibitemShut {NoStop}%
\bibitem [{\citenamefont {Horodecki}\ and\ \citenamefont
  {Piani}(2012)}]{horodecki2012}%
  \BibitemOpen
  \bibfield  {author} {\bibinfo {author} {\bibfnamefont {M.}~\bibnamefont
  {Horodecki}}\ and\ \bibinfo {author} {\bibfnamefont {M.}~\bibnamefont
  {Piani}},\ }\href {\doibase 10.1088/1751-8113/45/10/105306} {\bibfield
  {journal} {\bibinfo  {journal} {J. Phys. A: Math. Theor.}\ }\textbf {\bibinfo
  {volume} {45}},\ \bibinfo {pages} {105306} (\bibinfo {year}
  {2012})}\BibitemShut {NoStop}%
\bibitem [{\citenamefont {Shadman}\ \emph {et~al.}(2012)\citenamefont
  {Shadman}, \citenamefont {Kampermann}, \citenamefont {Bru\ss{}},\ and\
  \citenamefont {Macchiavello}}]{shadman2012}%
  \BibitemOpen
  \bibfield  {author} {\bibinfo {author} {\bibfnamefont {Z.}~\bibnamefont
  {Shadman}}, \bibinfo {author} {\bibfnamefont {H.}~\bibnamefont {Kampermann}},
  \bibinfo {author} {\bibfnamefont {D.}~\bibnamefont {Bru\ss{}}}, \ and\
  \bibinfo {author} {\bibfnamefont {C.}~\bibnamefont {Macchiavello}},\ }\href
  {\doibase 10.1103/PhysRevA.85.052306} {\bibfield  {journal} {\bibinfo
  {journal} {Phys. Rev. A}\ }\textbf {\bibinfo {volume} {85}},\ \bibinfo
  {pages} {052306} (\bibinfo {year} {2012})}\BibitemShut {NoStop}%
\bibitem [{\citenamefont {Das}\ \emph {et~al.}(2014)\citenamefont {Das},
  \citenamefont {Prabhu}, \citenamefont {Sen(De)},\ and\ \citenamefont
  {Sen}}]{das2014}%
  \BibitemOpen
  \bibfield  {author} {\bibinfo {author} {\bibfnamefont {T.}~\bibnamefont
  {Das}}, \bibinfo {author} {\bibfnamefont {R.}~\bibnamefont {Prabhu}},
  \bibinfo {author} {\bibfnamefont {A.}~\bibnamefont {Sen(De)}}, \ and\
  \bibinfo {author} {\bibfnamefont {U.}~\bibnamefont {Sen}},\ }\href {\doibase
  10.1103/PhysRevA.90.022319} {\bibfield  {journal} {\bibinfo  {journal} {Phys.
  Rev. A}\ }\textbf {\bibinfo {volume} {90}},\ \bibinfo {pages} {022319}
  (\bibinfo {year} {2014})}\BibitemShut {NoStop}%
\bibitem [{\citenamefont {Das}\ \emph {et~al.}(2015)\citenamefont {Das},
  \citenamefont {Prabhu}, \citenamefont {Sen(De)},\ and\ \citenamefont
  {Sen}}]{das2015}%
  \BibitemOpen
  \bibfield  {author} {\bibinfo {author} {\bibfnamefont {T.}~\bibnamefont
  {Das}}, \bibinfo {author} {\bibfnamefont {R.}~\bibnamefont {Prabhu}},
  \bibinfo {author} {\bibfnamefont {A.}~\bibnamefont {Sen(De)}}, \ and\
  \bibinfo {author} {\bibfnamefont {U.}~\bibnamefont {Sen}},\ }\href {\doibase
  10.1103/PhysRevA.92.052330} {\bibfield  {journal} {\bibinfo  {journal} {Phys.
  Rev. A}\ }\textbf {\bibinfo {volume} {92}},\ \bibinfo {pages} {052330}
  (\bibinfo {year} {2015})}\BibitemShut {NoStop}%
\bibitem [{\citenamefont {Ekert}(1991)}]{ekert1991}%
  \BibitemOpen
  \bibfield  {author} {\bibinfo {author} {\bibfnamefont {A.~K.}\ \bibnamefont
  {Ekert}},\ }\href {\doibase 10.1103/PhysRevLett.67.661} {\bibfield  {journal}
  {\bibinfo  {journal} {Phys. Rev. Lett.}\ }\textbf {\bibinfo {volume} {67}},\
  \bibinfo {pages} {661} (\bibinfo {year} {1991})}\BibitemShut {NoStop}%
\bibitem [{\citenamefont {Jennewein}\ \emph {et~al.}(2000)\citenamefont
  {Jennewein}, \citenamefont {Simon}, \citenamefont {Weihs}, \citenamefont
  {Weinfurter},\ and\ \citenamefont {Zeilinger}}]{jennewein2000}%
  \BibitemOpen
  \bibfield  {author} {\bibinfo {author} {\bibfnamefont {T.}~\bibnamefont
  {Jennewein}}, \bibinfo {author} {\bibfnamefont {C.}~\bibnamefont {Simon}},
  \bibinfo {author} {\bibfnamefont {G.}~\bibnamefont {Weihs}}, \bibinfo
  {author} {\bibfnamefont {H.}~\bibnamefont {Weinfurter}}, \ and\ \bibinfo
  {author} {\bibfnamefont {A.}~\bibnamefont {Zeilinger}},\ }\href {\doibase
  10.1103/PhysRevLett.84.4729} {\bibfield  {journal} {\bibinfo  {journal}
  {Phys. Rev. Lett.}\ }\textbf {\bibinfo {volume} {84}},\ \bibinfo {pages}
  {4729} (\bibinfo {year} {2000})}\BibitemShut {NoStop}%
\bibitem [{\citenamefont {Gisin}\ \emph {et~al.}(2002)\citenamefont {Gisin},
  \citenamefont {Ribordy}, \citenamefont {Tittel},\ and\ \citenamefont
  {Zbinden}}]{gisin2002}%
  \BibitemOpen
  \bibfield  {author} {\bibinfo {author} {\bibfnamefont {N.}~\bibnamefont
  {Gisin}}, \bibinfo {author} {\bibfnamefont {G.}~\bibnamefont {Ribordy}},
  \bibinfo {author} {\bibfnamefont {W.}~\bibnamefont {Tittel}}, \ and\ \bibinfo
  {author} {\bibfnamefont {H.}~\bibnamefont {Zbinden}},\ }\href {\doibase
  10.1103/RevModPhys.74.145} {\bibfield  {journal} {\bibinfo  {journal} {Rev.
  Mod. Phys.}\ }\textbf {\bibinfo {volume} {74}},\ \bibinfo {pages} {145}
  (\bibinfo {year} {2002})}\BibitemShut {NoStop}%
\bibitem [{\citenamefont {Pirandola}\ \emph {et~al.}(2019)\citenamefont
  {Pirandola}, \citenamefont {Andersen}, \citenamefont {Banchi}, \citenamefont
  {Berta}, \citenamefont {Bunandar}, \citenamefont {Colbeck}, \citenamefont
  {Englund}, \citenamefont {Gehring}, \citenamefont {Lupo}, \citenamefont
  {Ottaviani}, \citenamefont {Pereira}, \citenamefont {Razavi}, \citenamefont
  {Shaari}, \citenamefont {Tomamichel}, \citenamefont {Usenko}, \citenamefont
  {Vallone}, \citenamefont {Villoresi},\ and\ \citenamefont
  {Wallden}}]{pirandola2019}%
  \BibitemOpen
  \bibfield  {author} {\bibinfo {author} {\bibfnamefont {S.}~\bibnamefont
  {Pirandola}}, \bibinfo {author} {\bibfnamefont {U.~L.}\ \bibnamefont
  {Andersen}}, \bibinfo {author} {\bibfnamefont {L.}~\bibnamefont {Banchi}},
  \bibinfo {author} {\bibfnamefont {M.}~\bibnamefont {Berta}}, \bibinfo
  {author} {\bibfnamefont {D.}~\bibnamefont {Bunandar}}, \bibinfo {author}
  {\bibfnamefont {R.}~\bibnamefont {Colbeck}}, \bibinfo {author} {\bibfnamefont
  {D.}~\bibnamefont {Englund}}, \bibinfo {author} {\bibfnamefont
  {T.}~\bibnamefont {Gehring}}, \bibinfo {author} {\bibfnamefont
  {C.}~\bibnamefont {Lupo}}, \bibinfo {author} {\bibfnamefont {C.}~\bibnamefont
  {Ottaviani}}, \bibinfo {author} {\bibfnamefont {J.}~\bibnamefont {Pereira}},
  \bibinfo {author} {\bibfnamefont {M.}~\bibnamefont {Razavi}}, \bibinfo
  {author} {\bibfnamefont {J.~S.}\ \bibnamefont {Shaari}}, \bibinfo {author}
  {\bibfnamefont {M.}~\bibnamefont {Tomamichel}}, \bibinfo {author}
  {\bibfnamefont {V.~C.}\ \bibnamefont {Usenko}}, \bibinfo {author}
  {\bibfnamefont {G.}~\bibnamefont {Vallone}}, \bibinfo {author} {\bibfnamefont
  {P.}~\bibnamefont {Villoresi}}, \ and\ \bibinfo {author} {\bibfnamefont
  {P.}~\bibnamefont {Wallden}},\ }\href {https://arxiv.org/abs/1906.01645}
  {\bibfield  {journal} {\bibinfo  {journal} {arXiv:1906.01645}\ } (\bibinfo
  {year} {2019})}\BibitemShut {NoStop}%
\bibitem [{\citenamefont {Hillery}\ \emph {et~al.}(1999)\citenamefont
  {Hillery}, \citenamefont {Bu\ifmmode~\check{z}\else \v{z}\fi{}ek},\ and\
  \citenamefont {Berthiaume}}]{hillery1999}%
  \BibitemOpen
  \bibfield  {author} {\bibinfo {author} {\bibfnamefont {M.}~\bibnamefont
  {Hillery}}, \bibinfo {author} {\bibfnamefont {V.}~\bibnamefont
  {Bu\ifmmode~\check{z}\else \v{z}\fi{}ek}}, \ and\ \bibinfo {author}
  {\bibfnamefont {A.}~\bibnamefont {Berthiaume}},\ }\href {\doibase
  10.1103/PhysRevA.59.1829} {\bibfield  {journal} {\bibinfo  {journal} {Phys.
  Rev. A}\ }\textbf {\bibinfo {volume} {59}},\ \bibinfo {pages} {1829}
  (\bibinfo {year} {1999})}\BibitemShut {NoStop}%
\bibitem [{\citenamefont {Cleve}\ \emph {et~al.}(1999)\citenamefont {Cleve},
  \citenamefont {Gottesman},\ and\ \citenamefont {Lo}}]{cleve1999}%
  \BibitemOpen
  \bibfield  {author} {\bibinfo {author} {\bibfnamefont {R.}~\bibnamefont
  {Cleve}}, \bibinfo {author} {\bibfnamefont {D.}~\bibnamefont {Gottesman}}, \
  and\ \bibinfo {author} {\bibfnamefont {H.-K.}\ \bibnamefont {Lo}},\ }\href
  {\doibase 10.1103/PhysRevLett.83.648} {\bibfield  {journal} {\bibinfo
  {journal} {Phys. Rev. Lett.}\ }\textbf {\bibinfo {volume} {83}},\ \bibinfo
  {pages} {648} (\bibinfo {year} {1999})}\BibitemShut {NoStop}%
\bibitem [{\citenamefont {Karlsson}\ \emph {et~al.}(1999)\citenamefont
  {Karlsson}, \citenamefont {Koashi},\ and\ \citenamefont
  {Imoto}}]{karlsson1999}%
  \BibitemOpen
  \bibfield  {author} {\bibinfo {author} {\bibfnamefont {A.}~\bibnamefont
  {Karlsson}}, \bibinfo {author} {\bibfnamefont {M.}~\bibnamefont {Koashi}}, \
  and\ \bibinfo {author} {\bibfnamefont {N.}~\bibnamefont {Imoto}},\ }\href
  {\doibase 10.1103/PhysRevA.59.162} {\bibfield  {journal} {\bibinfo  {journal}
  {Phys. Rev. A}\ }\textbf {\bibinfo {volume} {59}},\ \bibinfo {pages} {162}
  (\bibinfo {year} {1999})}\BibitemShut {NoStop}%
\bibitem [{\citenamefont {Raussendorf}\ and\ \citenamefont
  {Briegel}(2001)}]{raussendorf2001}%
  \BibitemOpen
  \bibfield  {author} {\bibinfo {author} {\bibfnamefont {R.}~\bibnamefont
  {Raussendorf}}\ and\ \bibinfo {author} {\bibfnamefont {H.~J.}\ \bibnamefont
  {Briegel}},\ }\href {\doibase 10.1103/PhysRevLett.86.5188} {\bibfield
  {journal} {\bibinfo  {journal} {Phys. Rev. Lett.}\ }\textbf {\bibinfo
  {volume} {86}},\ \bibinfo {pages} {5188} (\bibinfo {year}
  {2001})}\BibitemShut {NoStop}%
\bibitem [{\citenamefont {Hein}\ \emph {et~al.}(2004)\citenamefont {Hein},
  \citenamefont {Eisert},\ and\ \citenamefont {Briegel}}]{hein2004}%
  \BibitemOpen
  \bibfield  {author} {\bibinfo {author} {\bibfnamefont {M.}~\bibnamefont
  {Hein}}, \bibinfo {author} {\bibfnamefont {J.}~\bibnamefont {Eisert}}, \ and\
  \bibinfo {author} {\bibfnamefont {H.~J.}\ \bibnamefont {Briegel}},\ }\href
  {\doibase 10.1103/PhysRevA.69.062311} {\bibfield  {journal} {\bibinfo
  {journal} {Phys. Rev. A}\ }\textbf {\bibinfo {volume} {69}},\ \bibinfo
  {pages} {062311} (\bibinfo {year} {2004})}\BibitemShut {NoStop}%
\bibitem [{\citenamefont {Hein}\ \emph {et~al.}(2006)\citenamefont {Hein},
  \citenamefont {D\"{u}r}, \citenamefont {Eisert}, \citenamefont {Raussendorf},
  \citenamefont {Van~den Nest},\ and\ \citenamefont {J.~Briegel}}]{hein2006}%
  \BibitemOpen
  \bibfield  {author} {\bibinfo {author} {\bibfnamefont {M.}~\bibnamefont
  {Hein}}, \bibinfo {author} {\bibfnamefont {W.}~\bibnamefont {D\"{u}r}},
  \bibinfo {author} {\bibfnamefont {J.}~\bibnamefont {Eisert}}, \bibinfo
  {author} {\bibfnamefont {R.}~\bibnamefont {Raussendorf}}, \bibinfo {author}
  {\bibfnamefont {M.}~\bibnamefont {Van~den Nest}}, \ and\ \bibinfo {author}
  {\bibfnamefont {H.}~\bibnamefont {J.~Briegel}},\ }in\ \href
  {https://arxiv.org/abs/quant-ph/0602096} {\emph {\bibinfo {booktitle}
  {Proceedings of the International School of Physics ``Enrico Fermi" on
  ``Quantum Computers, Algorithms and Chaos"}}}\ (\bibinfo {year} {2006})\
  \Eprint {http://arxiv.org/abs/arXiv:quant-ph/0602096}
  {arXiv:quant-ph/0602096} \BibitemShut {NoStop}%
\bibitem [{\citenamefont {Briegel}\ \emph {et~al.}(2009)\citenamefont
  {Briegel}, \citenamefont {Browne}, \citenamefont {D{\"u}r}, \citenamefont
  {Raussendorf},\ and\ \citenamefont {Van~den Nest}}]{briegel2009}%
  \BibitemOpen
  \bibfield  {author} {\bibinfo {author} {\bibfnamefont {H.~J.}\ \bibnamefont
  {Briegel}}, \bibinfo {author} {\bibfnamefont {D.~E.}\ \bibnamefont {Browne}},
  \bibinfo {author} {\bibfnamefont {W.}~\bibnamefont {D{\"u}r}}, \bibinfo
  {author} {\bibfnamefont {R.}~\bibnamefont {Raussendorf}}, \ and\ \bibinfo
  {author} {\bibfnamefont {M.}~\bibnamefont {Van~den Nest}},\ }\href
  {https://doi.org/10.1038/nphys1157} {\bibfield  {journal} {\bibinfo
  {journal} {Nat. Phys.}\ }\textbf {\bibinfo {volume} {5}},\ \bibinfo {pages}
  {19} (\bibinfo {year} {2009})}\BibitemShut {NoStop}%
\bibitem [{\citenamefont {Amico}\ \emph {et~al.}(2008)\citenamefont {Amico},
  \citenamefont {Fazio}, \citenamefont {Osterloh},\ and\ \citenamefont
  {Vedral}}]{amico2008}%
  \BibitemOpen
  \bibfield  {author} {\bibinfo {author} {\bibfnamefont {L.}~\bibnamefont
  {Amico}}, \bibinfo {author} {\bibfnamefont {R.}~\bibnamefont {Fazio}},
  \bibinfo {author} {\bibfnamefont {A.}~\bibnamefont {Osterloh}}, \ and\
  \bibinfo {author} {\bibfnamefont {V.}~\bibnamefont {Vedral}},\ }\href
  {\doibase 10.1103/RevModPhys.80.517} {\bibfield  {journal} {\bibinfo
  {journal} {Rev. Mod. Phys.}\ }\textbf {\bibinfo {volume} {80}},\ \bibinfo
  {pages} {517} (\bibinfo {year} {2008})}\BibitemShut {NoStop}%
\bibitem [{\citenamefont {Chiara}\ and\ \citenamefont
  {Sanpera}(2018)}]{chiara2018}%
  \BibitemOpen
  \bibfield  {author} {\bibinfo {author} {\bibfnamefont {G.~D.}\ \bibnamefont
  {Chiara}}\ and\ \bibinfo {author} {\bibfnamefont {A.}~\bibnamefont
  {Sanpera}},\ }\href {\doibase 10.1088/1361-6633/aabf61} {\bibfield  {journal}
  {\bibinfo  {journal} {Rep. Prog. Phys.}\ }\textbf {\bibinfo {volume} {81}},\
  \bibinfo {pages} {074002} (\bibinfo {year} {2018})}\BibitemShut {NoStop}%
\bibitem [{\citenamefont {Or\'us}(2008{\natexlab{a}})}]{orus2008}%
  \BibitemOpen
  \bibfield  {author} {\bibinfo {author} {\bibfnamefont {R.}~\bibnamefont
  {Or\'us}},\ }\href {\doibase 10.1103/PhysRevLett.100.130502} {\bibfield
  {journal} {\bibinfo  {journal} {Phys. Rev. Lett.}\ }\textbf {\bibinfo
  {volume} {100}},\ \bibinfo {pages} {130502} (\bibinfo {year}
  {2008}{\natexlab{a}})}\BibitemShut {NoStop}%
\bibitem [{\citenamefont {Wei}\ \emph {et~al.}(2005)\citenamefont {Wei},
  \citenamefont {Das}, \citenamefont {Mukhopadyay}, \citenamefont
  {Vishveshwara},\ and\ \citenamefont {Goldbart}}]{wei2005}%
  \BibitemOpen
  \bibfield  {author} {\bibinfo {author} {\bibfnamefont {T.-C.}\ \bibnamefont
  {Wei}}, \bibinfo {author} {\bibfnamefont {D.}~\bibnamefont {Das}}, \bibinfo
  {author} {\bibfnamefont {S.}~\bibnamefont {Mukhopadyay}}, \bibinfo {author}
  {\bibfnamefont {S.}~\bibnamefont {Vishveshwara}}, \ and\ \bibinfo {author}
  {\bibfnamefont {P.~M.}\ \bibnamefont {Goldbart}},\ }\href {\doibase
  10.1103/PhysRevA.71.060305} {\bibfield  {journal} {\bibinfo  {journal} {Phys.
  Rev. A}\ }\textbf {\bibinfo {volume} {71}},\ \bibinfo {pages} {060305}
  (\bibinfo {year} {2005})}\BibitemShut {NoStop}%
\bibitem [{\citenamefont {Or\'us}\ and\ \citenamefont {Wei}(2010)}]{orus2010}%
  \BibitemOpen
  \bibfield  {author} {\bibinfo {author} {\bibfnamefont {R.}~\bibnamefont
  {Or\'us}}\ and\ \bibinfo {author} {\bibfnamefont {T.-C.}\ \bibnamefont
  {Wei}},\ }\href {\doibase 10.1103/PhysRevB.82.155120} {\bibfield  {journal}
  {\bibinfo  {journal} {Phys. Rev. B}\ }\textbf {\bibinfo {volume} {82}},\
  \bibinfo {pages} {155120} (\bibinfo {year} {2010})}\BibitemShut {NoStop}%
\bibitem [{\citenamefont {Biswas}\ \emph {et~al.}(2014)\citenamefont {Biswas},
  \citenamefont {Prabhu}, \citenamefont {Sen(De)},\ and\ \citenamefont
  {Sen}}]{biswas2014}%
  \BibitemOpen
  \bibfield  {author} {\bibinfo {author} {\bibfnamefont {A.}~\bibnamefont
  {Biswas}}, \bibinfo {author} {\bibfnamefont {R.}~\bibnamefont {Prabhu}},
  \bibinfo {author} {\bibfnamefont {A.}~\bibnamefont {Sen(De)}}, \ and\
  \bibinfo {author} {\bibfnamefont {U.}~\bibnamefont {Sen}},\ }\href {\doibase
  10.1103/PhysRevA.90.032301} {\bibfield  {journal} {\bibinfo  {journal} {Phys.
  Rev. A}\ }\textbf {\bibinfo {volume} {90}},\ \bibinfo {pages} {032301}
  (\bibinfo {year} {2014})}\BibitemShut {NoStop}%
\bibitem [{\citenamefont {Or\'us}(2008{\natexlab{b}})}]{orus2008b}%
  \BibitemOpen
  \bibfield  {author} {\bibinfo {author} {\bibfnamefont {R.}~\bibnamefont
  {Or\'us}},\ }\href {\doibase 10.1103/PhysRevA.78.062332} {\bibfield
  {journal} {\bibinfo  {journal} {Phys. Rev. A}\ }\textbf {\bibinfo {volume}
  {78}},\ \bibinfo {pages} {062332} (\bibinfo {year}
  {2008}{\natexlab{b}})}\BibitemShut {NoStop}%
\bibitem [{\citenamefont {Dhar}\ \emph {et~al.}(2013)\citenamefont {Dhar},
  \citenamefont {Sen(De)},\ and\ \citenamefont {Sen}}]{dhar2013}%
  \BibitemOpen
  \bibfield  {author} {\bibinfo {author} {\bibfnamefont {H.~S.}\ \bibnamefont
  {Dhar}}, \bibinfo {author} {\bibfnamefont {A.}~\bibnamefont {Sen(De)}}, \
  and\ \bibinfo {author} {\bibfnamefont {U.}~\bibnamefont {Sen}},\ }\href
  {\doibase 10.1103/PhysRevLett.111.070501} {\bibfield  {journal} {\bibinfo
  {journal} {Phys. Rev. Lett.}\ }\textbf {\bibinfo {volume} {111}},\ \bibinfo
  {pages} {070501} (\bibinfo {year} {2013})}\BibitemShut {NoStop}%
\bibitem [{\citenamefont {Pezz\`e}\ \emph {et~al.}(2017)\citenamefont
  {Pezz\`e}, \citenamefont {Gabbrielli}, \citenamefont {Lepori},\ and\
  \citenamefont {Smerzi}}]{pezze2017}%
  \BibitemOpen
  \bibfield  {author} {\bibinfo {author} {\bibfnamefont {L.}~\bibnamefont
  {Pezz\`e}}, \bibinfo {author} {\bibfnamefont {M.}~\bibnamefont {Gabbrielli}},
  \bibinfo {author} {\bibfnamefont {L.}~\bibnamefont {Lepori}}, \ and\ \bibinfo
  {author} {\bibfnamefont {A.}~\bibnamefont {Smerzi}},\ }\href {\doibase
  10.1103/PhysRevLett.119.250401} {\bibfield  {journal} {\bibinfo  {journal}
  {Phys. Rev. Lett.}\ }\textbf {\bibinfo {volume} {119}},\ \bibinfo {pages}
  {250401} (\bibinfo {year} {2017})}\BibitemShut {NoStop}%
\bibitem [{\citenamefont {Mandel}\ \emph {et~al.}(2003)\citenamefont {Mandel},
  \citenamefont {Greiner}, \citenamefont {Widera}, \citenamefont {Rom},
  \citenamefont {Hansch},\ and\ \citenamefont {Bloch}}]{mandel2003}%
  \BibitemOpen
  \bibfield  {author} {\bibinfo {author} {\bibfnamefont {O.}~\bibnamefont
  {Mandel}}, \bibinfo {author} {\bibfnamefont {M.}~\bibnamefont {Greiner}},
  \bibinfo {author} {\bibfnamefont {A.}~\bibnamefont {Widera}}, \bibinfo
  {author} {\bibfnamefont {T.}~\bibnamefont {Rom}}, \bibinfo {author}
  {\bibfnamefont {T.~W.}\ \bibnamefont {Hansch}}, \ and\ \bibinfo {author}
  {\bibfnamefont {I.}~\bibnamefont {Bloch}},\ }\href {\doibase
  10.1038/nature02008} {\bibfield  {journal} {\bibinfo  {journal} {Nature}\
  }\textbf {\bibinfo {volume} {425}},\ \bibinfo {pages} {937} (\bibinfo {year}
  {2003})}\BibitemShut {NoStop}%
\bibitem [{\citenamefont {Leibfried}\ \emph {et~al.}(2005)\citenamefont
  {Leibfried}, \citenamefont {Knill}, \citenamefont {Seidelin}, \citenamefont
  {Britton}, \citenamefont {Blakestad}, \citenamefont {Chiaverini},
  \citenamefont {Hume}, \citenamefont {Itano}, \citenamefont {Jost},
  \citenamefont {Langer}, \citenamefont {Ozeri}, \citenamefont {Reichle},\ and\
  \citenamefont {Wineland}}]{leibfried2005}%
  \BibitemOpen
  \bibfield  {author} {\bibinfo {author} {\bibfnamefont {D.}~\bibnamefont
  {Leibfried}}, \bibinfo {author} {\bibfnamefont {E.}~\bibnamefont {Knill}},
  \bibinfo {author} {\bibfnamefont {S.}~\bibnamefont {Seidelin}}, \bibinfo
  {author} {\bibfnamefont {J.}~\bibnamefont {Britton}}, \bibinfo {author}
  {\bibfnamefont {R.~B.}\ \bibnamefont {Blakestad}}, \bibinfo {author}
  {\bibfnamefont {J.}~\bibnamefont {Chiaverini}}, \bibinfo {author}
  {\bibfnamefont {D.~B.}\ \bibnamefont {Hume}}, \bibinfo {author}
  {\bibfnamefont {W.~M.}\ \bibnamefont {Itano}}, \bibinfo {author}
  {\bibfnamefont {J.~D.}\ \bibnamefont {Jost}}, \bibinfo {author}
  {\bibfnamefont {C.}~\bibnamefont {Langer}}, \bibinfo {author} {\bibfnamefont
  {R.}~\bibnamefont {Ozeri}}, \bibinfo {author} {\bibfnamefont
  {R.}~\bibnamefont {Reichle}}, \ and\ \bibinfo {author} {\bibfnamefont
  {D.~J.}\ \bibnamefont {Wineland}},\ }\href {\doibase 10.1038/nature04251}
  {\bibfield  {journal} {\bibinfo  {journal} {Nature}\ }\textbf {\bibinfo
  {volume} {438}},\ \bibinfo {pages} {639} (\bibinfo {year}
  {2005})}\BibitemShut {NoStop}%
\bibitem [{\citenamefont {Monz}\ \emph {et~al.}(2011)\citenamefont {Monz},
  \citenamefont {Schindler}, \citenamefont {Barreiro}, \citenamefont {Chwalla},
  \citenamefont {Nigg}, \citenamefont {Coish}, \citenamefont {Harlander},
  \citenamefont {H\"ansel}, \citenamefont {Hennrich},\ and\ \citenamefont
  {Blatt}}]{monz2011}%
  \BibitemOpen
  \bibfield  {author} {\bibinfo {author} {\bibfnamefont {T.}~\bibnamefont
  {Monz}}, \bibinfo {author} {\bibfnamefont {P.}~\bibnamefont {Schindler}},
  \bibinfo {author} {\bibfnamefont {J.~T.}\ \bibnamefont {Barreiro}}, \bibinfo
  {author} {\bibfnamefont {M.}~\bibnamefont {Chwalla}}, \bibinfo {author}
  {\bibfnamefont {D.}~\bibnamefont {Nigg}}, \bibinfo {author} {\bibfnamefont
  {W.~A.}\ \bibnamefont {Coish}}, \bibinfo {author} {\bibfnamefont
  {M.}~\bibnamefont {Harlander}}, \bibinfo {author} {\bibfnamefont
  {W.}~\bibnamefont {H\"ansel}}, \bibinfo {author} {\bibfnamefont
  {M.}~\bibnamefont {Hennrich}}, \ and\ \bibinfo {author} {\bibfnamefont
  {R.}~\bibnamefont {Blatt}},\ }\href {\doibase 10.1103/PhysRevLett.106.130506}
  {\bibfield  {journal} {\bibinfo  {journal} {Phys. Rev. Lett.}\ }\textbf
  {\bibinfo {volume} {106}},\ \bibinfo {pages} {130506} (\bibinfo {year}
  {2011})}\BibitemShut {NoStop}%
\bibitem [{\citenamefont {Friis}\ \emph {et~al.}(2018)\citenamefont {Friis},
  \citenamefont {Marty}, \citenamefont {Maier}, \citenamefont {Hempel},
  \citenamefont {Holz\"apfel}, \citenamefont {Jurcevic}, \citenamefont
  {Plenio}, \citenamefont {Huber}, \citenamefont {Roos}, \citenamefont
  {Blatt},\ and\ \citenamefont {Lanyon}}]{friis2018}%
  \BibitemOpen
  \bibfield  {author} {\bibinfo {author} {\bibfnamefont {N.}~\bibnamefont
  {Friis}}, \bibinfo {author} {\bibfnamefont {O.}~\bibnamefont {Marty}},
  \bibinfo {author} {\bibfnamefont {C.}~\bibnamefont {Maier}}, \bibinfo
  {author} {\bibfnamefont {C.}~\bibnamefont {Hempel}}, \bibinfo {author}
  {\bibfnamefont {M.}~\bibnamefont {Holz\"apfel}}, \bibinfo {author}
  {\bibfnamefont {P.}~\bibnamefont {Jurcevic}}, \bibinfo {author}
  {\bibfnamefont {M.~B.}\ \bibnamefont {Plenio}}, \bibinfo {author}
  {\bibfnamefont {M.}~\bibnamefont {Huber}}, \bibinfo {author} {\bibfnamefont
  {C.}~\bibnamefont {Roos}}, \bibinfo {author} {\bibfnamefont {R.}~\bibnamefont
  {Blatt}}, \ and\ \bibinfo {author} {\bibfnamefont {B.}~\bibnamefont
  {Lanyon}},\ }\href {\doibase 10.1103/PhysRevX.8.021012} {\bibfield  {journal}
  {\bibinfo  {journal} {Phys. Rev. X}\ }\textbf {\bibinfo {volume} {8}},\
  \bibinfo {pages} {021012} (\bibinfo {year} {2018})}\BibitemShut {NoStop}%
\bibitem [{\citenamefont {Prevedel}\ \emph {et~al.}(2009)\citenamefont
  {Prevedel}, \citenamefont {Cronenberg}, \citenamefont {Tame}, \citenamefont
  {Paternostro}, \citenamefont {Walther}, \citenamefont {Kim},\ and\
  \citenamefont {Zeilinger}}]{prevedel2009}%
  \BibitemOpen
  \bibfield  {author} {\bibinfo {author} {\bibfnamefont {R.}~\bibnamefont
  {Prevedel}}, \bibinfo {author} {\bibfnamefont {G.}~\bibnamefont
  {Cronenberg}}, \bibinfo {author} {\bibfnamefont {M.~S.}\ \bibnamefont
  {Tame}}, \bibinfo {author} {\bibfnamefont {M.}~\bibnamefont {Paternostro}},
  \bibinfo {author} {\bibfnamefont {P.}~\bibnamefont {Walther}}, \bibinfo
  {author} {\bibfnamefont {M.~S.}\ \bibnamefont {Kim}}, \ and\ \bibinfo
  {author} {\bibfnamefont {A.}~\bibnamefont {Zeilinger}},\ }\href {\doibase
  10.1103/PhysRevLett.103.020503} {\bibfield  {journal} {\bibinfo  {journal}
  {Phys. Rev. Lett.}\ }\textbf {\bibinfo {volume} {103}},\ \bibinfo {pages}
  {020503} (\bibinfo {year} {2009})}\BibitemShut {NoStop}%
\bibitem [{\citenamefont {Gao}\ \emph {et~al.}(2010)\citenamefont {Gao},
  \citenamefont {Lu}, \citenamefont {Yao}, \citenamefont {Xu}, \citenamefont
  {G{\"{u}}hne}, \citenamefont {Goebel}, \citenamefont {Chen}, \citenamefont
  {Peng}, \citenamefont {Chen},\ and\ \citenamefont {Pan}}]{gao2010}%
  \BibitemOpen
  \bibfield  {author} {\bibinfo {author} {\bibfnamefont {W.-B.}\ \bibnamefont
  {Gao}}, \bibinfo {author} {\bibfnamefont {C.-Y.}\ \bibnamefont {Lu}},
  \bibinfo {author} {\bibfnamefont {X.-C.}\ \bibnamefont {Yao}}, \bibinfo
  {author} {\bibfnamefont {P.}~\bibnamefont {Xu}}, \bibinfo {author}
  {\bibfnamefont {O.}~\bibnamefont {G{\"{u}}hne}}, \bibinfo {author}
  {\bibfnamefont {A.}~\bibnamefont {Goebel}}, \bibinfo {author} {\bibfnamefont
  {Y.-A.}\ \bibnamefont {Chen}}, \bibinfo {author} {\bibfnamefont {C.-Z.}\
  \bibnamefont {Peng}}, \bibinfo {author} {\bibfnamefont {Z.-B.}\ \bibnamefont
  {Chen}}, \ and\ \bibinfo {author} {\bibfnamefont {J.-W.}\ \bibnamefont
  {Pan}},\ }\href {https://doi.org/10.1038/nphys1603 http://10.0.4.14/nphys1603
  https://www.nature.com/articles/nphys1603{\#}supplementary-information}
  {\bibfield  {journal} {\bibinfo  {journal} {Nature Physics}\ }\textbf
  {\bibinfo {volume} {6}},\ \bibinfo {pages} {331} (\bibinfo {year}
  {2010})}\BibitemShut {NoStop}%
\bibitem [{\citenamefont {Yao}\ \emph {et~al.}(2012)\citenamefont {Yao},
  \citenamefont {Wang}, \citenamefont {Xu}, \citenamefont {Lu}, \citenamefont
  {Pan}, \citenamefont {Bao}, \citenamefont {Peng}, \citenamefont {Lu},
  \citenamefont {Chen},\ and\ \citenamefont {Pan}}]{yao2012}%
  \BibitemOpen
  \bibfield  {author} {\bibinfo {author} {\bibfnamefont {X.-C.}\ \bibnamefont
  {Yao}}, \bibinfo {author} {\bibfnamefont {T.-X.}\ \bibnamefont {Wang}},
  \bibinfo {author} {\bibfnamefont {P.}~\bibnamefont {Xu}}, \bibinfo {author}
  {\bibfnamefont {H.}~\bibnamefont {Lu}}, \bibinfo {author} {\bibfnamefont
  {G.-S.}\ \bibnamefont {Pan}}, \bibinfo {author} {\bibfnamefont {X.-H.}\
  \bibnamefont {Bao}}, \bibinfo {author} {\bibfnamefont {C.-Z.}\ \bibnamefont
  {Peng}}, \bibinfo {author} {\bibfnamefont {C.-Y.}\ \bibnamefont {Lu}},
  \bibinfo {author} {\bibfnamefont {Y.-A.}\ \bibnamefont {Chen}}, \ and\
  \bibinfo {author} {\bibfnamefont {J.-W.}\ \bibnamefont {Pan}},\ }\href
  {https://doi.org/10.1038/nphoton.2011.354 http://10.0.4.14/nphoton.2011.354
  https://www.nature.com/articles/nphoton.2011.354{\#}supplementary-information}
  {\bibfield  {journal} {\bibinfo  {journal} {Nature Photonics}\ }\textbf
  {\bibinfo {volume} {6}},\ \bibinfo {pages} {225} (\bibinfo {year}
  {2012})}\BibitemShut {NoStop}%
\bibitem [{\citenamefont {Wang}\ \emph {et~al.}(2018)\citenamefont {Wang},
  \citenamefont {Luo}, \citenamefont {Huang}, \citenamefont {Chen},
  \citenamefont {Su}, \citenamefont {Liu}, \citenamefont {Chen}, \citenamefont
  {Li}, \citenamefont {Fang}, \citenamefont {Jiang}, \citenamefont {Zhang},
  \citenamefont {Li}, \citenamefont {Liu}, \citenamefont {Lu},\ and\
  \citenamefont {Pan}}]{wang2018}%
  \BibitemOpen
  \bibfield  {author} {\bibinfo {author} {\bibfnamefont {X.-L.}\ \bibnamefont
  {Wang}}, \bibinfo {author} {\bibfnamefont {Y.-H.}\ \bibnamefont {Luo}},
  \bibinfo {author} {\bibfnamefont {H.-L.}\ \bibnamefont {Huang}}, \bibinfo
  {author} {\bibfnamefont {M.-C.}\ \bibnamefont {Chen}}, \bibinfo {author}
  {\bibfnamefont {Z.-E.}\ \bibnamefont {Su}}, \bibinfo {author} {\bibfnamefont
  {C.}~\bibnamefont {Liu}}, \bibinfo {author} {\bibfnamefont {C.}~\bibnamefont
  {Chen}}, \bibinfo {author} {\bibfnamefont {W.}~\bibnamefont {Li}}, \bibinfo
  {author} {\bibfnamefont {Y.-Q.}\ \bibnamefont {Fang}}, \bibinfo {author}
  {\bibfnamefont {X.}~\bibnamefont {Jiang}}, \bibinfo {author} {\bibfnamefont
  {J.}~\bibnamefont {Zhang}}, \bibinfo {author} {\bibfnamefont
  {L.}~\bibnamefont {Li}}, \bibinfo {author} {\bibfnamefont {N.-L.}\
  \bibnamefont {Liu}}, \bibinfo {author} {\bibfnamefont {C.-Y.}\ \bibnamefont
  {Lu}}, \ and\ \bibinfo {author} {\bibfnamefont {J.-W.}\ \bibnamefont {Pan}},\
  }\href {\doibase 10.1103/PhysRevLett.120.260502} {\bibfield  {journal}
  {\bibinfo  {journal} {Phys. Rev. Lett.}\ }\textbf {\bibinfo {volume} {120}},\
  \bibinfo {pages} {260502} (\bibinfo {year} {2018})}\BibitemShut {NoStop}%
\bibitem [{\citenamefont {Barends}\ \emph {et~al.}(2014)\citenamefont
  {Barends}, \citenamefont {Kelly}, \citenamefont {Megrant}, \citenamefont
  {Veitia}, \citenamefont {Sank}, \citenamefont {Jeffrey}, \citenamefont
  {White}, \citenamefont {Mutus}, \citenamefont {Fowler}, \citenamefont
  {Campbell}, \citenamefont {Chen}, \citenamefont {Chen}, \citenamefont
  {Chiaro}, \citenamefont {Dunsworth}, \citenamefont {Neill}, \citenamefont
  {O'Malley}, \citenamefont {Roushan}, \citenamefont {Vainsencher},
  \citenamefont {Wenner}, \citenamefont {Korotkov}, \citenamefont {Cleland},\
  and\ \citenamefont {Martinis}}]{barends2014}%
  \BibitemOpen
  \bibfield  {author} {\bibinfo {author} {\bibfnamefont {R.}~\bibnamefont
  {Barends}}, \bibinfo {author} {\bibfnamefont {J.}~\bibnamefont {Kelly}},
  \bibinfo {author} {\bibfnamefont {A.}~\bibnamefont {Megrant}}, \bibinfo
  {author} {\bibfnamefont {A.}~\bibnamefont {Veitia}}, \bibinfo {author}
  {\bibfnamefont {D.}~\bibnamefont {Sank}}, \bibinfo {author} {\bibfnamefont
  {E.}~\bibnamefont {Jeffrey}}, \bibinfo {author} {\bibfnamefont {T.~C.}\
  \bibnamefont {White}}, \bibinfo {author} {\bibfnamefont {J.}~\bibnamefont
  {Mutus}}, \bibinfo {author} {\bibfnamefont {A.~G.}\ \bibnamefont {Fowler}},
  \bibinfo {author} {\bibfnamefont {B.}~\bibnamefont {Campbell}}, \bibinfo
  {author} {\bibfnamefont {Y.}~\bibnamefont {Chen}}, \bibinfo {author}
  {\bibfnamefont {Z.}~\bibnamefont {Chen}}, \bibinfo {author} {\bibfnamefont
  {B.}~\bibnamefont {Chiaro}}, \bibinfo {author} {\bibfnamefont
  {A.}~\bibnamefont {Dunsworth}}, \bibinfo {author} {\bibfnamefont
  {C.}~\bibnamefont {Neill}}, \bibinfo {author} {\bibfnamefont
  {P.}~\bibnamefont {O'Malley}}, \bibinfo {author} {\bibfnamefont
  {P.}~\bibnamefont {Roushan}}, \bibinfo {author} {\bibfnamefont
  {A.}~\bibnamefont {Vainsencher}}, \bibinfo {author} {\bibfnamefont
  {J.}~\bibnamefont {Wenner}}, \bibinfo {author} {\bibfnamefont {A.~N.}\
  \bibnamefont {Korotkov}}, \bibinfo {author} {\bibfnamefont {A.~N.}\
  \bibnamefont {Cleland}}, \ and\ \bibinfo {author} {\bibfnamefont {J.~M.}\
  \bibnamefont {Martinis}},\ }\href {https://doi.org/10.1038/nature13171}
  {\bibfield  {journal} {\bibinfo  {journal} {Nature}\ }\textbf {\bibinfo
  {volume} {508}},\ \bibinfo {pages} {500} (\bibinfo {year}
  {2014})}\BibitemShut {NoStop}%
\bibitem [{\citenamefont {Gong}\ \emph {et~al.}(2019)\citenamefont {Gong},
  \citenamefont {Chen}, \citenamefont {Zheng}, \citenamefont {Wang},
  \citenamefont {Zha}, \citenamefont {Deng}, \citenamefont {Yan}, \citenamefont
  {Rong}, \citenamefont {Wu}, \citenamefont {Li}, \citenamefont {Chen},
  \citenamefont {Zhao}, \citenamefont {Liang}, \citenamefont {Lin},
  \citenamefont {Xu}, \citenamefont {Guo}, \citenamefont {Sun}, \citenamefont
  {Castellano}, \citenamefont {Wang}, \citenamefont {Peng}, \citenamefont {Lu},
  \citenamefont {Zhu},\ and\ \citenamefont {Pan}}]{gong2019}%
  \BibitemOpen
  \bibfield  {author} {\bibinfo {author} {\bibfnamefont {M.}~\bibnamefont
  {Gong}}, \bibinfo {author} {\bibfnamefont {M.-C.}\ \bibnamefont {Chen}},
  \bibinfo {author} {\bibfnamefont {Y.}~\bibnamefont {Zheng}}, \bibinfo
  {author} {\bibfnamefont {S.}~\bibnamefont {Wang}}, \bibinfo {author}
  {\bibfnamefont {C.}~\bibnamefont {Zha}}, \bibinfo {author} {\bibfnamefont
  {H.}~\bibnamefont {Deng}}, \bibinfo {author} {\bibfnamefont {Z.}~\bibnamefont
  {Yan}}, \bibinfo {author} {\bibfnamefont {H.}~\bibnamefont {Rong}}, \bibinfo
  {author} {\bibfnamefont {Y.}~\bibnamefont {Wu}}, \bibinfo {author}
  {\bibfnamefont {S.}~\bibnamefont {Li}}, \bibinfo {author} {\bibfnamefont
  {F.}~\bibnamefont {Chen}}, \bibinfo {author} {\bibfnamefont {Y.}~\bibnamefont
  {Zhao}}, \bibinfo {author} {\bibfnamefont {F.}~\bibnamefont {Liang}},
  \bibinfo {author} {\bibfnamefont {J.}~\bibnamefont {Lin}}, \bibinfo {author}
  {\bibfnamefont {Y.}~\bibnamefont {Xu}}, \bibinfo {author} {\bibfnamefont
  {C.}~\bibnamefont {Guo}}, \bibinfo {author} {\bibfnamefont {L.}~\bibnamefont
  {Sun}}, \bibinfo {author} {\bibfnamefont {A.~D.}\ \bibnamefont {Castellano}},
  \bibinfo {author} {\bibfnamefont {H.}~\bibnamefont {Wang}}, \bibinfo {author}
  {\bibfnamefont {C.}~\bibnamefont {Peng}}, \bibinfo {author} {\bibfnamefont
  {C.-Y.}\ \bibnamefont {Lu}}, \bibinfo {author} {\bibfnamefont
  {X.}~\bibnamefont {Zhu}}, \ and\ \bibinfo {author} {\bibfnamefont {J.-W.}\
  \bibnamefont {Pan}},\ }\href {\doibase 10.1103/PhysRevLett.122.110501}
  {\bibfield  {journal} {\bibinfo  {journal} {Phys. Rev. Lett.}\ }\textbf
  {\bibinfo {volume} {122}},\ \bibinfo {pages} {110501} (\bibinfo {year}
  {2019})}\BibitemShut {NoStop}%
\bibitem [{\citenamefont {Negrevergne}\ \emph {et~al.}(2006)\citenamefont
  {Negrevergne}, \citenamefont {Mahesh}, \citenamefont {Ryan}, \citenamefont
  {Ditty}, \citenamefont {Cyr-Racine}, \citenamefont {Power}, \citenamefont
  {Boulant}, \citenamefont {Havel}, \citenamefont {Cory},\ and\ \citenamefont
  {Laflamme}}]{negrevergne2006}%
  \BibitemOpen
  \bibfield  {author} {\bibinfo {author} {\bibfnamefont {C.}~\bibnamefont
  {Negrevergne}}, \bibinfo {author} {\bibfnamefont {T.~S.}\ \bibnamefont
  {Mahesh}}, \bibinfo {author} {\bibfnamefont {C.~A.}\ \bibnamefont {Ryan}},
  \bibinfo {author} {\bibfnamefont {M.}~\bibnamefont {Ditty}}, \bibinfo
  {author} {\bibfnamefont {F.}~\bibnamefont {Cyr-Racine}}, \bibinfo {author}
  {\bibfnamefont {W.}~\bibnamefont {Power}}, \bibinfo {author} {\bibfnamefont
  {N.}~\bibnamefont {Boulant}}, \bibinfo {author} {\bibfnamefont
  {T.}~\bibnamefont {Havel}}, \bibinfo {author} {\bibfnamefont {D.~G.}\
  \bibnamefont {Cory}}, \ and\ \bibinfo {author} {\bibfnamefont
  {R.}~\bibnamefont {Laflamme}},\ }\href {\doibase
  10.1103/PhysRevLett.96.170501} {\bibfield  {journal} {\bibinfo  {journal}
  {Phys. Rev. Lett.}\ }\textbf {\bibinfo {volume} {96}},\ \bibinfo {pages}
  {170501} (\bibinfo {year} {2006})}\BibitemShut {NoStop}%
\bibitem [{\citenamefont {Bradley}\ \emph {et~al.}(2019)\citenamefont
  {Bradley}, \citenamefont {Randall}, \citenamefont {Abobeih}, \citenamefont
  {Berrevoets}, \citenamefont {Degen}, \citenamefont {Bakker}, \citenamefont
  {Markham}, \citenamefont {Twitchen},\ and\ \citenamefont
  {Taminiau}}]{bradley2019}%
  \BibitemOpen
  \bibfield  {author} {\bibinfo {author} {\bibfnamefont {C.~E.}\ \bibnamefont
  {Bradley}}, \bibinfo {author} {\bibfnamefont {J.}~\bibnamefont {Randall}},
  \bibinfo {author} {\bibfnamefont {M.~H.}\ \bibnamefont {Abobeih}}, \bibinfo
  {author} {\bibfnamefont {R.~C.}\ \bibnamefont {Berrevoets}}, \bibinfo
  {author} {\bibfnamefont {M.~J.}\ \bibnamefont {Degen}}, \bibinfo {author}
  {\bibfnamefont {M.~A.}\ \bibnamefont {Bakker}}, \bibinfo {author}
  {\bibfnamefont {M.}~\bibnamefont {Markham}}, \bibinfo {author} {\bibfnamefont
  {D.~J.}\ \bibnamefont {Twitchen}}, \ and\ \bibinfo {author} {\bibfnamefont
  {T.~H.}\ \bibnamefont {Taminiau}},\ }\href {https://arxiv.org/abs/1905.02094}
  {\bibfield  {journal} {\bibinfo  {journal} {arXiv:1905.02094}\ } (\bibinfo
  {year} {2019})}\BibitemShut {NoStop}%
\bibitem [{\citenamefont {Bengtsson}\ and\ \citenamefont
  {\.{Z}yczkowski}(2006)}]{bengtsson2006}%
  \BibitemOpen
  \bibfield  {author} {\bibinfo {author} {\bibfnamefont {I.}~\bibnamefont
  {Bengtsson}}\ and\ \bibinfo {author} {\bibfnamefont {K.}~\bibnamefont
  {\.{Z}yczkowski}},\ }\href@noop {} {\emph {\bibinfo {title} {Geometry of
  quantum states: An introduction to quantum entanglement}}}\ (\bibinfo
  {publisher} {Cambridge University Press},\ \bibinfo {year}
  {2006})\BibitemShut {NoStop}%
\bibitem [{\citenamefont {Hayden}\ \emph {et~al.}(2006)\citenamefont {Hayden},
  \citenamefont {Leung},\ and\ \citenamefont {Winter}}]{hayden2006}%
  \BibitemOpen
  \bibfield  {author} {\bibinfo {author} {\bibfnamefont {P.}~\bibnamefont
  {Hayden}}, \bibinfo {author} {\bibfnamefont {D.}~\bibnamefont {Leung}}, \
  and\ \bibinfo {author} {\bibfnamefont {A.}~\bibnamefont {Winter}},\ }\href
  {\doibase 10.1007/s00220-006-1535-6} {\bibfield  {journal} {\bibinfo
  {journal} {Commun. Math. Phys.}\ }\textbf {\bibinfo {volume} {265}},\
  \bibinfo {pages} {95} (\bibinfo {year} {2006})}\BibitemShut {NoStop}%
\bibitem [{\citenamefont {Kendon}\ \emph {et~al.}(2002)\citenamefont {Kendon},
  \citenamefont {\.Zyczkowski},\ and\ \citenamefont {Munro}}]{kendon2002}%
  \BibitemOpen
  \bibfield  {author} {\bibinfo {author} {\bibfnamefont {V.~M.}\ \bibnamefont
  {Kendon}}, \bibinfo {author} {\bibfnamefont {K.}~\bibnamefont
  {\.Zyczkowski}}, \ and\ \bibinfo {author} {\bibfnamefont {W.~J.}\
  \bibnamefont {Munro}},\ }\href {\doibase 10.1103/PhysRevA.66.062310}
  {\bibfield  {journal} {\bibinfo  {journal} {Phys. Rev. A}\ }\textbf {\bibinfo
  {volume} {66}},\ \bibinfo {pages} {062310} (\bibinfo {year}
  {2002})}\BibitemShut {NoStop}%
\bibitem [{\citenamefont {Enriquez}\ \emph {et~al.}(2018)\citenamefont
  {Enriquez}, \citenamefont {Delgado},\ and\ \citenamefont
  {\.{Z}yczkowski}}]{enriquez2018}%
  \BibitemOpen
  \bibfield  {author} {\bibinfo {author} {\bibfnamefont {M.}~\bibnamefont
  {Enriquez}}, \bibinfo {author} {\bibfnamefont {F.}~\bibnamefont {Delgado}}, \
  and\ \bibinfo {author} {\bibfnamefont {K.}~\bibnamefont {\.{Z}yczkowski}},\
  }\href {https://arxiv.org/abs/1809.00642} {\bibfield  {journal} {\bibinfo
  {journal} {arXiv:1809.00642}\ } (\bibinfo {year} {2018})}\BibitemShut
  {NoStop}%
\bibitem [{\citenamefont {Klobus}\ \emph {et~al.}(2019)\citenamefont {Klobus},
  \citenamefont {Burchardt}, \citenamefont {Kolodziejski}, \citenamefont
  {Pandit}, \citenamefont {Vertesi}, \citenamefont {\.{Z}yczkowski},\ and\
  \citenamefont {Laskowski}}]{klobus2019}%
  \BibitemOpen
  \bibfield  {author} {\bibinfo {author} {\bibfnamefont {W.}~\bibnamefont
  {Klobus}}, \bibinfo {author} {\bibfnamefont {A.}~\bibnamefont {Burchardt}},
  \bibinfo {author} {\bibfnamefont {A.}~\bibnamefont {Kolodziejski}}, \bibinfo
  {author} {\bibfnamefont {M.}~\bibnamefont {Pandit}}, \bibinfo {author}
  {\bibfnamefont {T.}~\bibnamefont {Vertesi}}, \bibinfo {author} {\bibfnamefont
  {K.}~\bibnamefont {\.{Z}yczkowski}}, \ and\ \bibinfo {author} {\bibfnamefont
  {W.}~\bibnamefont {Laskowski}},\ }\href {https://arxiv.org/abs/1906.01311}
  {\bibfield  {journal} {\bibinfo  {journal} {arXiv:1906.01311}\ } (\bibinfo
  {year} {2019})}\BibitemShut {NoStop}%
\bibitem [{\citenamefont {Shimony}(1995)}]{shimony1995}%
  \BibitemOpen
  \bibfield  {author} {\bibinfo {author} {\bibfnamefont {A.}~\bibnamefont
  {Shimony}},\ }\href {\doibase 10.1111/j.1749-6632.1995.tb39008.x} {\bibfield
  {journal} {\bibinfo  {journal} {Annals of the New York Academy of Sciences}\
  }\textbf {\bibinfo {volume} {755}},\ \bibinfo {pages} {675} (\bibinfo {year}
  {1995})}\BibitemShut {NoStop}%
\bibitem [{\citenamefont {Barnum}\ and\ \citenamefont
  {Linden}(2001)}]{barnum2001}%
  \BibitemOpen
  \bibfield  {author} {\bibinfo {author} {\bibfnamefont {H.}~\bibnamefont
  {Barnum}}\ and\ \bibinfo {author} {\bibfnamefont {N.}~\bibnamefont
  {Linden}},\ }\href {\doibase 10.1088/0305-4470/34/35/305} {\bibfield
  {journal} {\bibinfo  {journal} {Journal of Physics A: Mathematical and
  General}\ }\textbf {\bibinfo {volume} {34}},\ \bibinfo {pages} {6787}
  (\bibinfo {year} {2001})}\BibitemShut {NoStop}%
\bibitem [{\citenamefont {Wei}\ and\ \citenamefont {Goldbart}(2003)}]{wei2003}%
  \BibitemOpen
  \bibfield  {author} {\bibinfo {author} {\bibfnamefont {T.-C.}\ \bibnamefont
  {Wei}}\ and\ \bibinfo {author} {\bibfnamefont {P.~M.}\ \bibnamefont
  {Goldbart}},\ }\href {\doibase 10.1103/PhysRevA.68.042307} {\bibfield
  {journal} {\bibinfo  {journal} {Phys. Rev. A}\ }\textbf {\bibinfo {volume}
  {68}},\ \bibinfo {pages} {042307} (\bibinfo {year} {2003})}\BibitemShut
  {NoStop}%
\bibitem [{\citenamefont {Gross}\ \emph {et~al.}(2009)\citenamefont {Gross},
  \citenamefont {Flammia},\ and\ \citenamefont {Eisert}}]{gross2009}%
  \BibitemOpen
  \bibfield  {author} {\bibinfo {author} {\bibfnamefont {D.}~\bibnamefont
  {Gross}}, \bibinfo {author} {\bibfnamefont {S.~T.}\ \bibnamefont {Flammia}},
  \ and\ \bibinfo {author} {\bibfnamefont {J.}~\bibnamefont {Eisert}},\ }\href
  {\doibase 10.1103/PhysRevLett.102.190501} {\bibfield  {journal} {\bibinfo
  {journal} {Phys. Rev. Lett.}\ }\textbf {\bibinfo {volume} {102}},\ \bibinfo
  {pages} {190501} (\bibinfo {year} {2009})}\BibitemShut {NoStop}%
\bibitem [{\citenamefont {Bremner}\ \emph {et~al.}(2009)\citenamefont
  {Bremner}, \citenamefont {Mora},\ and\ \citenamefont {Winter}}]{bremner2009}%
  \BibitemOpen
  \bibfield  {author} {\bibinfo {author} {\bibfnamefont {M.~J.}\ \bibnamefont
  {Bremner}}, \bibinfo {author} {\bibfnamefont {C.}~\bibnamefont {Mora}}, \
  and\ \bibinfo {author} {\bibfnamefont {A.}~\bibnamefont {Winter}},\ }\href
  {\doibase 10.1103/PhysRevLett.102.190502} {\bibfield  {journal} {\bibinfo
  {journal} {Phys. Rev. Lett.}\ }\textbf {\bibinfo {volume} {102}},\ \bibinfo
  {pages} {190502} (\bibinfo {year} {2009})}\BibitemShut {NoStop}%
\bibitem [{\citenamefont {Richard}\ and\ \citenamefont
  {Noah}(2003)}]{jozsa2003}%
  \BibitemOpen
  \bibfield  {author} {\bibinfo {author} {\bibfnamefont {J.}~\bibnamefont
  {Richard}}\ and\ \bibinfo {author} {\bibfnamefont {L.}~\bibnamefont {Noah}},\
  }\href {\doibase 10.1098/rspa.2002.1097} {\bibfield  {journal} {\bibinfo
  {journal} {Proc. R. Soc. Lond. A}\ }\textbf {\bibinfo {volume} {459}}
  (\bibinfo {year} {2003}),\ 10.1098/rspa.2002.1097}\BibitemShut {NoStop}%
\bibitem [{\citenamefont {Coffman}\ \emph {et~al.}(2000)\citenamefont
  {Coffman}, \citenamefont {Kundu},\ and\ \citenamefont
  {Wootters}}]{coffman2003}%
  \BibitemOpen
  \bibfield  {author} {\bibinfo {author} {\bibfnamefont {V.}~\bibnamefont
  {Coffman}}, \bibinfo {author} {\bibfnamefont {J.}~\bibnamefont {Kundu}}, \
  and\ \bibinfo {author} {\bibfnamefont {W.~K.}\ \bibnamefont {Wootters}},\
  }\href {\doibase 10.1103/PhysRevA.61.052306} {\bibfield  {journal} {\bibinfo
  {journal} {Phys. Rev. A}\ }\textbf {\bibinfo {volume} {61}},\ \bibinfo
  {pages} {052306} (\bibinfo {year} {2000})}\BibitemShut {NoStop}%
\bibitem [{\citenamefont {Dhar}\ \emph {et~al.}(2017)\citenamefont {Dhar},
  \citenamefont {Pal}, \citenamefont {Rakshit}, \citenamefont {Sen(De)},\ and\
  \citenamefont {Sen}}]{dhar2017}%
  \BibitemOpen
  \bibfield  {author} {\bibinfo {author} {\bibfnamefont {H.~S.}\ \bibnamefont
  {Dhar}}, \bibinfo {author} {\bibfnamefont {A.~K.}\ \bibnamefont {Pal}},
  \bibinfo {author} {\bibfnamefont {D.}~\bibnamefont {Rakshit}}, \bibinfo
  {author} {\bibfnamefont {A.}~\bibnamefont {Sen(De)}}, \ and\ \bibinfo
  {author} {\bibfnamefont {U.}~\bibnamefont {Sen}},\ }in\ \href
  {https://arxiv.org/abs/1610.01069} {\emph {\bibinfo {booktitle} {Lectures on
  General Quantum Correlations and their Applications}}},\ \bibinfo {editor}
  {edited by\ \bibinfo {editor} {\bibfnamefont {F.~F.}\ \bibnamefont
  {Fanchini}}, \bibinfo {editor} {\bibfnamefont {D.~O.~S.}\ \bibnamefont
  {Pinto}}, \ and\ \bibinfo {editor} {\bibfnamefont {G.}~\bibnamefont
  {Adesso}}}\ (\bibinfo  {publisher} {Springer},\ \bibinfo {year} {2017})\
  \Eprint {http://arxiv.org/abs/arXiv:1610.01069} {arXiv:1610.01069}
  \BibitemShut {NoStop}%
\bibitem [{\citenamefont {Modi}\ \emph {et~al.}(2012)\citenamefont {Modi},
  \citenamefont {Brodutch}, \citenamefont {Cable}, \citenamefont {Paterek},\
  and\ \citenamefont {Vedral}}]{modi2012}%
  \BibitemOpen
  \bibfield  {author} {\bibinfo {author} {\bibfnamefont {K.}~\bibnamefont
  {Modi}}, \bibinfo {author} {\bibfnamefont {A.}~\bibnamefont {Brodutch}},
  \bibinfo {author} {\bibfnamefont {H.}~\bibnamefont {Cable}}, \bibinfo
  {author} {\bibfnamefont {T.}~\bibnamefont {Paterek}}, \ and\ \bibinfo
  {author} {\bibfnamefont {V.}~\bibnamefont {Vedral}},\ }\href {\doibase
  10.1103/RevModPhys.84.1655} {\bibfield  {journal} {\bibinfo  {journal} {Rev.
  Mod. Phys.}\ }\textbf {\bibinfo {volume} {84}},\ \bibinfo {pages} {1655}
  (\bibinfo {year} {2012})}\BibitemShut {NoStop}%
\bibitem [{\citenamefont {Bera}\ \emph {et~al.}(2017)\citenamefont {Bera},
  \citenamefont {Das}, \citenamefont {Sadhukhan}, \citenamefont {SinghaRoy},
  \citenamefont {Sen(De)},\ and\ \citenamefont {Sen}}]{bera2017}%
  \BibitemOpen
  \bibfield  {author} {\bibinfo {author} {\bibfnamefont {A.}~\bibnamefont
  {Bera}}, \bibinfo {author} {\bibfnamefont {T.}~\bibnamefont {Das}}, \bibinfo
  {author} {\bibfnamefont {D.}~\bibnamefont {Sadhukhan}}, \bibinfo {author}
  {\bibfnamefont {S.}~\bibnamefont {SinghaRoy}}, \bibinfo {author}
  {\bibfnamefont {A.}~\bibnamefont {Sen(De)}}, \ and\ \bibinfo {author}
  {\bibfnamefont {U.}~\bibnamefont {Sen}},\ }\href {\doibase
  10.1088/1361-6633/aa872f} {\bibfield  {journal} {\bibinfo  {journal} {Rep.
  Prog. Phys.}\ }\textbf {\bibinfo {volume} {81}},\ \bibinfo {pages} {024001}
  (\bibinfo {year} {2017})}\BibitemShut {NoStop}%
\bibitem [{\citenamefont {Rethinasamy}\ \emph {et~al.}(2019)\citenamefont
  {Rethinasamy}, \citenamefont {Roy}, \citenamefont {Chanda}, \citenamefont
  {Sen(De)},\ and\ \citenamefont {Sen}}]{rethinasamy2019}%
  \BibitemOpen
  \bibfield  {author} {\bibinfo {author} {\bibfnamefont {S.}~\bibnamefont
  {Rethinasamy}}, \bibinfo {author} {\bibfnamefont {S.}~\bibnamefont {Roy}},
  \bibinfo {author} {\bibfnamefont {T.}~\bibnamefont {Chanda}}, \bibinfo
  {author} {\bibfnamefont {A.}~\bibnamefont {Sen(De)}}, \ and\ \bibinfo
  {author} {\bibfnamefont {U.}~\bibnamefont {Sen}},\ }\href {\doibase
  10.1103/PhysRevA.99.042302} {\bibfield  {journal} {\bibinfo  {journal} {Phys.
  Rev. A}\ }\textbf {\bibinfo {volume} {99}},\ \bibinfo {pages} {042302}
  (\bibinfo {year} {2019})}\BibitemShut {NoStop}%
\bibitem [{\citenamefont {Verstraete}\ \emph
  {et~al.}(2004{\natexlab{a}})\citenamefont {Verstraete}, \citenamefont
  {Popp},\ and\ \citenamefont {Cirac}}]{verstraete2004}%
  \BibitemOpen
  \bibfield  {author} {\bibinfo {author} {\bibfnamefont {F.}~\bibnamefont
  {Verstraete}}, \bibinfo {author} {\bibfnamefont {M.}~\bibnamefont {Popp}}, \
  and\ \bibinfo {author} {\bibfnamefont {J.~I.}\ \bibnamefont {Cirac}},\ }\href
  {\doibase 10.1103/PhysRevLett.92.027901} {\bibfield  {journal} {\bibinfo
  {journal} {Phys. Rev. Lett.}\ }\textbf {\bibinfo {volume} {92}},\ \bibinfo
  {pages} {027901} (\bibinfo {year} {2004}{\natexlab{a}})}\BibitemShut
  {NoStop}%
\bibitem [{\citenamefont {Verstraete}\ \emph
  {et~al.}(2004{\natexlab{b}})\citenamefont {Verstraete}, \citenamefont
  {Mart\'{\i}n-Delgado},\ and\ \citenamefont {Cirac}}]{verstraete2004a}%
  \BibitemOpen
  \bibfield  {author} {\bibinfo {author} {\bibfnamefont {F.}~\bibnamefont
  {Verstraete}}, \bibinfo {author} {\bibfnamefont {M.~A.}\ \bibnamefont
  {Mart\'{\i}n-Delgado}}, \ and\ \bibinfo {author} {\bibfnamefont {J.~I.}\
  \bibnamefont {Cirac}},\ }\href {\doibase 10.1103/PhysRevLett.92.087201}
  {\bibfield  {journal} {\bibinfo  {journal} {Phys. Rev. Lett.}\ }\textbf
  {\bibinfo {volume} {92}},\ \bibinfo {pages} {087201} (\bibinfo {year}
  {2004}{\natexlab{b}})}\BibitemShut {NoStop}%
\bibitem [{\citenamefont {Popp}\ \emph {et~al.}(2005)\citenamefont {Popp},
  \citenamefont {Verstraete}, \citenamefont {Mart\'{\i}n-Delgado},\ and\
  \citenamefont {Cirac}}]{popp2005}%
  \BibitemOpen
  \bibfield  {author} {\bibinfo {author} {\bibfnamefont {M.}~\bibnamefont
  {Popp}}, \bibinfo {author} {\bibfnamefont {F.}~\bibnamefont {Verstraete}},
  \bibinfo {author} {\bibfnamefont {M.~A.}\ \bibnamefont
  {Mart\'{\i}n-Delgado}}, \ and\ \bibinfo {author} {\bibfnamefont {J.~I.}\
  \bibnamefont {Cirac}},\ }\href {\doibase 10.1103/PhysRevA.71.042306}
  {\bibfield  {journal} {\bibinfo  {journal} {Phys. Rev. A}\ }\textbf {\bibinfo
  {volume} {71}},\ \bibinfo {pages} {042306} (\bibinfo {year}
  {2005})}\BibitemShut {NoStop}%
\bibitem [{\citenamefont {Jin}\ and\ \citenamefont {Korepin}(2004)}]{jin2004}%
  \BibitemOpen
  \bibfield  {author} {\bibinfo {author} {\bibfnamefont {B.-Q.}\ \bibnamefont
  {Jin}}\ and\ \bibinfo {author} {\bibfnamefont {V.~E.}\ \bibnamefont
  {Korepin}},\ }\href {\doibase 10.1103/PhysRevA.69.062314} {\bibfield
  {journal} {\bibinfo  {journal} {Phys. Rev. A}\ }\textbf {\bibinfo {volume}
  {69}},\ \bibinfo {pages} {062314} (\bibinfo {year} {2004})}\BibitemShut
  {NoStop}%
\bibitem [{\citenamefont {DiVincenzo}\ \emph {et~al.}(1998)\citenamefont
  {DiVincenzo}, \citenamefont {Fuchs}, \citenamefont {Mabuchi}, \citenamefont
  {Smolin}, \citenamefont {Thapliyal},\ and\ \citenamefont
  {Uhlmann}}]{divincenzo1998}%
  \BibitemOpen
  \bibfield  {author} {\bibinfo {author} {\bibfnamefont {D.~P.}\ \bibnamefont
  {DiVincenzo}}, \bibinfo {author} {\bibfnamefont {C.~A.}\ \bibnamefont
  {Fuchs}}, \bibinfo {author} {\bibfnamefont {H.}~\bibnamefont {Mabuchi}},
  \bibinfo {author} {\bibfnamefont {J.~A.}\ \bibnamefont {Smolin}}, \bibinfo
  {author} {\bibfnamefont {A.}~\bibnamefont {Thapliyal}}, \ and\ \bibinfo
  {author} {\bibfnamefont {A.}~\bibnamefont {Uhlmann}},\ }\href
  {https://arxiv.org/abs/quant-ph/9803033} {\bibfield  {journal} {\bibinfo
  {journal} {arXiv:quant-ph/9803033}\ } (\bibinfo {year} {1998})}\BibitemShut
  {NoStop}%
\bibitem [{\citenamefont {Smolin}\ \emph {et~al.}(2005)\citenamefont {Smolin},
  \citenamefont {Verstraete},\ and\ \citenamefont {Winter}}]{smolin2005}%
  \BibitemOpen
  \bibfield  {author} {\bibinfo {author} {\bibfnamefont {J.~A.}\ \bibnamefont
  {Smolin}}, \bibinfo {author} {\bibfnamefont {F.}~\bibnamefont {Verstraete}},
  \ and\ \bibinfo {author} {\bibfnamefont {A.}~\bibnamefont {Winter}},\ }\href
  {\doibase 10.1103/PhysRevA.72.052317} {\bibfield  {journal} {\bibinfo
  {journal} {Phys. Rev. A}\ }\textbf {\bibinfo {volume} {72}},\ \bibinfo
  {pages} {052317} (\bibinfo {year} {2005})}\BibitemShut {NoStop}%
\bibitem [{\citenamefont {Streltsov}\ \emph {et~al.}(2015)\citenamefont
  {Streltsov}, \citenamefont {Lee},\ and\ \citenamefont
  {Adesso}}]{streltsov2015}%
  \BibitemOpen
  \bibfield  {author} {\bibinfo {author} {\bibfnamefont {A.}~\bibnamefont
  {Streltsov}}, \bibinfo {author} {\bibfnamefont {S.}~\bibnamefont {Lee}}, \
  and\ \bibinfo {author} {\bibfnamefont {G.}~\bibnamefont {Adesso}},\ }\href
  {\doibase 10.1103/PhysRevLett.115.030505} {\bibfield  {journal} {\bibinfo
  {journal} {Phys. Rev. Lett.}\ }\textbf {\bibinfo {volume} {115}},\ \bibinfo
  {pages} {030505} (\bibinfo {year} {2015})}\BibitemShut {NoStop}%
\bibitem [{\citenamefont {Amaro}\ \emph {et~al.}(2018)\citenamefont {Amaro},
  \citenamefont {M\"{u}ller},\ and\ \citenamefont {Pal}}]{amaro2018}%
  \BibitemOpen
  \bibfield  {author} {\bibinfo {author} {\bibfnamefont {D.}~\bibnamefont
  {Amaro}}, \bibinfo {author} {\bibfnamefont {M.}~\bibnamefont {M\"{u}ller}}, \
  and\ \bibinfo {author} {\bibfnamefont {A.~K.}\ \bibnamefont {Pal}},\ }\href
  {\doibase 10.1088/1367-2630/aac485} {\bibfield  {journal} {\bibinfo
  {journal} {New J. Phys.}\ }\textbf {\bibinfo {volume} {20}},\ \bibinfo
  {pages} {063017} (\bibinfo {year} {2018})}\BibitemShut {NoStop}%
\bibitem [{\citenamefont {Amaro}\ \emph {et~al.}(2019)\citenamefont {Amaro},
  \citenamefont {M\"{u}ller},\ and\ \citenamefont {Pal}}]{amaro2019}%
  \BibitemOpen
  \bibfield  {author} {\bibinfo {author} {\bibfnamefont {D.}~\bibnamefont
  {Amaro}}, \bibinfo {author} {\bibfnamefont {M.}~\bibnamefont {M\"{u}ller}}, \
  and\ \bibinfo {author} {\bibfnamefont {A.~K.}\ \bibnamefont {Pal}},\ }\href
  {https://arxiv.org/abs/1907.13161} {\bibfield  {journal} {\bibinfo  {journal}
  {arXiv:1907.13161}\ } (\bibinfo {year} {2019})}\BibitemShut {NoStop}%
\bibitem [{\citenamefont {Ac{\'i}n}\ \emph {et~al.}(2007)\citenamefont
  {Ac{\'i}n}, \citenamefont {Cirac},\ and\ \citenamefont
  {Lewenstein}}]{acin2007}%
  \BibitemOpen
  \bibfield  {author} {\bibinfo {author} {\bibfnamefont {A.}~\bibnamefont
  {Ac{\'i}n}}, \bibinfo {author} {\bibfnamefont {J.~I.}\ \bibnamefont {Cirac}},
  \ and\ \bibinfo {author} {\bibfnamefont {M.}~\bibnamefont {Lewenstein}},\
  }\href {https://doi.org/10.1038/nphys549} {\bibfield  {journal} {\bibinfo
  {journal} {Nat. Phys.}\ }\textbf {\bibinfo {volume} {3}},\ \bibinfo {pages}
  {256} (\bibinfo {year} {2007})}\BibitemShut {NoStop}%
\bibitem [{\citenamefont {Sadhukhan}\ \emph {et~al.}(2017)\citenamefont
  {Sadhukhan}, \citenamefont {SinnghaRoy}, \citenamefont {Pal}, \citenamefont
  {Rakshit}, \citenamefont {Sen(De)},\ and\ \citenamefont
  {Sen}}]{sadhukhan2017}%
  \BibitemOpen
  \bibfield  {author} {\bibinfo {author} {\bibfnamefont {D.}~\bibnamefont
  {Sadhukhan}}, \bibinfo {author} {\bibfnamefont {S.}~\bibnamefont
  {SinnghaRoy}}, \bibinfo {author} {\bibfnamefont {A.~K.}\ \bibnamefont {Pal}},
  \bibinfo {author} {\bibfnamefont {D.}~\bibnamefont {Rakshit}}, \bibinfo
  {author} {\bibfnamefont {A.}~\bibnamefont {Sen(De)}}, \ and\ \bibinfo
  {author} {\bibfnamefont {U.}~\bibnamefont {Sen}},\ }\href {\doibase
  10.1103/PhysRevA.95.022301} {\bibfield  {journal} {\bibinfo  {journal} {Phys.
  Rev. A}\ }\textbf {\bibinfo {volume} {95}},\ \bibinfo {pages} {022301}
  (\bibinfo {year} {2017})}\BibitemShut {NoStop}%
\bibitem [{\citenamefont {Gour}\ and\ \citenamefont
  {Spekkens}(2006)}]{gour2006}%
  \BibitemOpen
  \bibfield  {author} {\bibinfo {author} {\bibfnamefont {G.}~\bibnamefont
  {Gour}}\ and\ \bibinfo {author} {\bibfnamefont {R.~W.}\ \bibnamefont
  {Spekkens}},\ }\href {\doibase 10.1103/PhysRevA.73.062331} {\bibfield
  {journal} {\bibinfo  {journal} {Phys. Rev. A}\ }\textbf {\bibinfo {volume}
  {73}},\ \bibinfo {pages} {062331} (\bibinfo {year} {2006})}\BibitemShut
  {NoStop}%
\bibitem [{\citenamefont {Wootters}(1998)}]{wootters1998}%
  \BibitemOpen
  \bibfield  {author} {\bibinfo {author} {\bibfnamefont {W.~K.}\ \bibnamefont
  {Wootters}},\ }\href {\doibase 10.1103/PhysRevLett.80.2245} {\bibfield
  {journal} {\bibinfo  {journal} {Phys. Rev. Lett.}\ }\textbf {\bibinfo
  {volume} {80}},\ \bibinfo {pages} {2245} (\bibinfo {year}
  {1998})}\BibitemShut {NoStop}%
\bibitem [{\citenamefont {Wootters}(2001)}]{wootters2001}%
  \BibitemOpen
  \bibfield  {author} {\bibinfo {author} {\bibfnamefont {W.~K.}\ \bibnamefont
  {Wootters}},\ }\href {\doibase 10.1103/PhysRevLett.80.2245} {\bibfield
  {journal} {\bibinfo  {journal} {Quant. Info. Comput.}\ }\textbf {\bibinfo
  {volume} {1}},\ \bibinfo {pages} {27} (\bibinfo {year} {2001})}\BibitemShut
  {NoStop}%
\bibitem [{\citenamefont {Vidal}\ and\ \citenamefont
  {Werner}(2002)}]{vidal2002}%
  \BibitemOpen
  \bibfield  {author} {\bibinfo {author} {\bibfnamefont {G.}~\bibnamefont
  {Vidal}}\ and\ \bibinfo {author} {\bibfnamefont {R.~F.}\ \bibnamefont
  {Werner}},\ }\href {\doibase 10.1103/PhysRevA.65.032314} {\bibfield
  {journal} {\bibinfo  {journal} {Phys. Rev. A}\ }\textbf {\bibinfo {volume}
  {65}},\ \bibinfo {pages} {032314} (\bibinfo {year} {2002})}\BibitemShut
  {NoStop}%
\bibitem [{\citenamefont {Nielsen}\ and\ \citenamefont
  {Chuang}(2010)}]{nielsen2010}%
  \BibitemOpen
  \bibfield  {author} {\bibinfo {author} {\bibfnamefont {M.~A.}\ \bibnamefont
  {Nielsen}}\ and\ \bibinfo {author} {\bibfnamefont {I.~L.}\ \bibnamefont
  {Chuang}},\ }\href@noop {} {\emph {\bibinfo {title} {Quantum Computation and
  Quantum Information}}}\ (\bibinfo  {publisher} {Cambridge University Press},\
  \bibinfo {year} {2010})\BibitemShut {NoStop}%
\bibitem [{\citenamefont {Venuti}\ and\ \citenamefont
  {Roncaglia}(2005)}]{venuti2005}%
  \BibitemOpen
  \bibfield  {author} {\bibinfo {author} {\bibfnamefont {L.~C.}\ \bibnamefont
  {Venuti}}\ and\ \bibinfo {author} {\bibfnamefont {M.}~\bibnamefont
  {Roncaglia}},\ }\href {\doibase 10.1103/PhysRevLett.94.207207} {\bibfield
  {journal} {\bibinfo  {journal} {Phys. Rev. Lett.}\ }\textbf {\bibinfo
  {volume} {94}},\ \bibinfo {pages} {207207} (\bibinfo {year}
  {2005})}\BibitemShut {NoStop}%
\bibitem [{\citenamefont {Holevo}\ and\ \citenamefont
  {Giovannetti}(2012)}]{holevo2012}%
  \BibitemOpen
  \bibfield  {author} {\bibinfo {author} {\bibfnamefont {A.~S.}\ \bibnamefont
  {Holevo}}\ and\ \bibinfo {author} {\bibfnamefont {V.}~\bibnamefont
  {Giovannetti}},\ }\href {\doibase 10.1088/0034-4885/75/4/046001} {\bibfield
  {journal} {\bibinfo  {journal} {Rep. Prog. Phys.}\ }\textbf {\bibinfo
  {volume} {75}},\ \bibinfo {pages} {046001} (\bibinfo {year}
  {2012})}\BibitemShut {NoStop}%
\bibitem [{\citenamefont {D\"ur}\ \emph {et~al.}(2000)\citenamefont {D\"ur},
  \citenamefont {Vidal},\ and\ \citenamefont {Cirac}}]{dur2000}%
  \BibitemOpen
  \bibfield  {author} {\bibinfo {author} {\bibfnamefont {W.}~\bibnamefont
  {D\"ur}}, \bibinfo {author} {\bibfnamefont {G.}~\bibnamefont {Vidal}}, \ and\
  \bibinfo {author} {\bibfnamefont {J.~I.}\ \bibnamefont {Cirac}},\ }\href
  {\doibase 10.1103/PhysRevA.62.062314} {\bibfield  {journal} {\bibinfo
  {journal} {Phys. Rev. A}\ }\textbf {\bibinfo {volume} {62}},\ \bibinfo
  {pages} {062314} (\bibinfo {year} {2000})}\BibitemShut {NoStop}%
\bibitem [{\citenamefont {Bulmer}(1965)}]{bulmer1965}%
  \BibitemOpen
  \bibfield  {author} {\bibinfo {author} {\bibfnamefont {M.~G.}\ \bibnamefont
  {Bulmer}},\ }\href@noop {} {\emph {\bibinfo {title} {Principles of
  Statistics}}}\ (\bibinfo  {publisher} {Dover Publications},\ \bibinfo {year}
  {1965})\BibitemShut {NoStop}%
\bibitem [{\citenamefont {Ozols}(2009)}]{ozols2009}%
  \BibitemOpen
  \bibfield  {author} {\bibinfo {author} {\bibfnamefont {M.}~\bibnamefont
  {Ozols}},\ }\href {http://home.lu.lv/~sd20008/papers/essays.html} {\bibfield
  {journal} {\bibinfo  {journal} {Essay on generation of random unitary
  matrices}\ } (\bibinfo {year} {2009})}\BibitemShut {NoStop}%
\bibitem [{\citenamefont {Sen(De)}\ \emph {et~al.}(2003)\citenamefont
  {Sen(De)}, \citenamefont {Sen}, \citenamefont {Wie\ifmmode~\acute{s}\else
  \'{s}\fi{}niak}, \citenamefont {Kaszlikowski},\ and\ \citenamefont
  {\ifmmode~\dot{Z}\else \.{Z}\fi{}ukowski}}]{sende2003}%
  \BibitemOpen
  \bibfield  {author} {\bibinfo {author} {\bibfnamefont {A.}~\bibnamefont
  {Sen(De)}}, \bibinfo {author} {\bibfnamefont {U.}~\bibnamefont {Sen}},
  \bibinfo {author} {\bibfnamefont {M.}~\bibnamefont
  {Wie\ifmmode~\acute{s}\else \'{s}\fi{}niak}}, \bibinfo {author}
  {\bibfnamefont {D.}~\bibnamefont {Kaszlikowski}}, \ and\ \bibinfo {author}
  {\bibfnamefont {M.}~\bibnamefont {\ifmmode~\dot{Z}\else \.{Z}\fi{}ukowski}},\
  }\href {\doibase 10.1103/PhysRevA.68.062306} {\bibfield  {journal} {\bibinfo
  {journal} {Phys. Rev. A}\ }\textbf {\bibinfo {volume} {68}},\ \bibinfo
  {pages} {062306} (\bibinfo {year} {2003})}\BibitemShut {NoStop}%
\bibitem [{\citenamefont {Kaszlikowski}\ \emph {et~al.}(2008)\citenamefont
  {Kaszlikowski}, \citenamefont {Sen(De)}, \citenamefont {Sen}, \citenamefont
  {Vedral},\ and\ \citenamefont {Winter}}]{kaszlikowski2008}%
  \BibitemOpen
  \bibfield  {author} {\bibinfo {author} {\bibfnamefont {D.}~\bibnamefont
  {Kaszlikowski}}, \bibinfo {author} {\bibfnamefont {A.}~\bibnamefont
  {Sen(De)}}, \bibinfo {author} {\bibfnamefont {U.}~\bibnamefont {Sen}},
  \bibinfo {author} {\bibfnamefont {V.}~\bibnamefont {Vedral}}, \ and\ \bibinfo
  {author} {\bibfnamefont {A.}~\bibnamefont {Winter}},\ }\href {\doibase
  10.1103/PhysRevLett.101.070502} {\bibfield  {journal} {\bibinfo  {journal}
  {Phys. Rev. Lett.}\ }\textbf {\bibinfo {volume} {101}},\ \bibinfo {pages}
  {070502} (\bibinfo {year} {2008})}\BibitemShut {NoStop}%
\bibitem [{\citenamefont {Barnea}\ \emph {et~al.}(2015)\citenamefont {Barnea},
  \citenamefont {P\"utz}, \citenamefont {Brask}, \citenamefont {Brunner},
  \citenamefont {Gisin},\ and\ \citenamefont {Liang}}]{barnea2015}%
  \BibitemOpen
  \bibfield  {author} {\bibinfo {author} {\bibfnamefont {T.~J.}\ \bibnamefont
  {Barnea}}, \bibinfo {author} {\bibfnamefont {G.}~\bibnamefont {P\"utz}},
  \bibinfo {author} {\bibfnamefont {J.~B.}\ \bibnamefont {Brask}}, \bibinfo
  {author} {\bibfnamefont {N.}~\bibnamefont {Brunner}}, \bibinfo {author}
  {\bibfnamefont {N.}~\bibnamefont {Gisin}}, \ and\ \bibinfo {author}
  {\bibfnamefont {Y.-C.}\ \bibnamefont {Liang}},\ }\href {\doibase
  10.1103/PhysRevA.91.032108} {\bibfield  {journal} {\bibinfo  {journal} {Phys.
  Rev. A}\ }\textbf {\bibinfo {volume} {91}},\ \bibinfo {pages} {032108}
  (\bibinfo {year} {2015})}\BibitemShut {NoStop}%
\bibitem [{\citenamefont {Laskowski}\ \emph {et~al.}(2015)\citenamefont
  {Laskowski}, \citenamefont {V{\'{e}}rtesi},\ and\ \citenamefont
  {Wie{\'{s}}niak}}]{laskowski2015}%
  \BibitemOpen
  \bibfield  {author} {\bibinfo {author} {\bibfnamefont {W.}~\bibnamefont
  {Laskowski}}, \bibinfo {author} {\bibfnamefont {T.}~\bibnamefont
  {V{\'{e}}rtesi}}, \ and\ \bibinfo {author} {\bibfnamefont {M.}~\bibnamefont
  {Wie{\'{s}}niak}},\ }\href {\doibase 10.1088/1751-8113/48/46/465301}
  {\bibfield  {journal} {\bibinfo  {journal} {J. Phys. A: Math. Theor.}\
  }\textbf {\bibinfo {volume} {48}},\ \bibinfo {pages} {465301} (\bibinfo
  {year} {2015})}\BibitemShut {NoStop}%
\bibitem [{\citenamefont {Roy}\ \emph {et~al.}(2018)\citenamefont {Roy},
  \citenamefont {Chanda}, \citenamefont {Das}, \citenamefont {Sen(De)},\ and\
  \citenamefont {Sen}}]{roy2018}%
  \BibitemOpen
  \bibfield  {author} {\bibinfo {author} {\bibfnamefont {S.}~\bibnamefont
  {Roy}}, \bibinfo {author} {\bibfnamefont {T.}~\bibnamefont {Chanda}},
  \bibinfo {author} {\bibfnamefont {T.}~\bibnamefont {Das}}, \bibinfo {author}
  {\bibfnamefont {A.}~\bibnamefont {Sen(De)}}, \ and\ \bibinfo {author}
  {\bibfnamefont {U.}~\bibnamefont {Sen}},\ }\href {\doibase
  10.1103/PhysRevA.91.032108} {\bibfield  {journal} {\bibinfo  {journal} {Phys.
  Lett. A}\ }\textbf {\bibinfo {volume} {382}},\ \bibinfo {pages} {1709}
  (\bibinfo {year} {2018})}\BibitemShut {NoStop}%
\bibitem [{\citenamefont {Giorgi}(2011)}]{giorgi2011}%
  \BibitemOpen
  \bibfield  {author} {\bibinfo {author} {\bibfnamefont {G.~L.}\ \bibnamefont
  {Giorgi}},\ }\href {\doibase 10.1103/PhysRevA.84.054301} {\bibfield
  {journal} {\bibinfo  {journal} {Phys. Rev. A}\ }\textbf {\bibinfo {volume}
  {84}},\ \bibinfo {pages} {054301} (\bibinfo {year} {2011})}\BibitemShut
  {NoStop}%
\bibitem [{\citenamefont {Prabhu}\ \emph {et~al.}(2012)\citenamefont {Prabhu},
  \citenamefont {Pati}, \citenamefont {Sen(De)},\ and\ \citenamefont
  {Sen}}]{prabhu2012}%
  \BibitemOpen
  \bibfield  {author} {\bibinfo {author} {\bibfnamefont {R.}~\bibnamefont
  {Prabhu}}, \bibinfo {author} {\bibfnamefont {A.~K.}\ \bibnamefont {Pati}},
  \bibinfo {author} {\bibfnamefont {A.}~\bibnamefont {Sen(De)}}, \ and\
  \bibinfo {author} {\bibfnamefont {U.}~\bibnamefont {Sen}},\ }\href {\doibase
  10.1103/PhysRevA.85.040102} {\bibfield  {journal} {\bibinfo  {journal} {Phys.
  Rev. A}\ }\textbf {\bibinfo {volume} {85}},\ \bibinfo {pages} {040102}
  (\bibinfo {year} {2012})}\BibitemShut {NoStop}%
\bibitem [{\citenamefont {Dicke}(1954)}]{dicke1954}%
  \BibitemOpen
  \bibfield  {author} {\bibinfo {author} {\bibfnamefont {R.~H.}\ \bibnamefont
  {Dicke}},\ }\href {\doibase 10.1103/PhysRev.93.99} {\bibfield  {journal}
  {\bibinfo  {journal} {Phys. Rev.}\ }\textbf {\bibinfo {volume} {93}},\
  \bibinfo {pages} {99} (\bibinfo {year} {1954})}\BibitemShut {NoStop}%
\bibitem [{\citenamefont {Kumar}\ \emph {et~al.}(2017)\citenamefont {Kumar},
  \citenamefont {Dhar}, \citenamefont {Prabhu}, \citenamefont {Sen(De)},\ and\
  \citenamefont {Sen}}]{kumar2017}%
  \BibitemOpen
  \bibfield  {author} {\bibinfo {author} {\bibfnamefont {A.}~\bibnamefont
  {Kumar}}, \bibinfo {author} {\bibfnamefont {H.~S.}\ \bibnamefont {Dhar}},
  \bibinfo {author} {\bibfnamefont {R.}~\bibnamefont {Prabhu}}, \bibinfo
  {author} {\bibfnamefont {A.}~\bibnamefont {Sen(De)}}, \ and\ \bibinfo
  {author} {\bibfnamefont {U.}~\bibnamefont {Sen}},\ }\href {\doibase
  https://doi.org/10.1016/j.physleta.2017.03.026} {\bibfield  {journal}
  {\bibinfo  {journal} {Phys. Lett. A}\ }\textbf {\bibinfo {volume} {381}},\
  \bibinfo {pages} {1701 } (\bibinfo {year} {2017})}\BibitemShut {NoStop}%
\bibitem [{\citenamefont {Bergmann}\ and\ \citenamefont
  {G\"{u}hne}(2013)}]{bergmann2013}%
  \BibitemOpen
  \bibfield  {author} {\bibinfo {author} {\bibfnamefont {M.}~\bibnamefont
  {Bergmann}}\ and\ \bibinfo {author} {\bibfnamefont {O.}~\bibnamefont
  {G\"{u}hne}},\ }\href {\doibase 10.1088/1751-8113/46/38/385304} {\bibfield
  {journal} {\bibinfo  {journal} {J. Phys. A: Math. Theor.}\ }\textbf {\bibinfo
  {volume} {46}},\ \bibinfo {pages} {385304} (\bibinfo {year}
  {2013})}\BibitemShut {NoStop}%
\bibitem [{\citenamefont {L\"ucke}\ \emph {et~al.}(2014)\citenamefont
  {L\"ucke}, \citenamefont {Peise}, \citenamefont {Vitagliano}, \citenamefont
  {Arlt}, \citenamefont {Santos}, \citenamefont {T\'oth},\ and\ \citenamefont
  {Klempt}}]{lucke2014}%
  \BibitemOpen
  \bibfield  {author} {\bibinfo {author} {\bibfnamefont {B.}~\bibnamefont
  {L\"ucke}}, \bibinfo {author} {\bibfnamefont {J.}~\bibnamefont {Peise}},
  \bibinfo {author} {\bibfnamefont {G.}~\bibnamefont {Vitagliano}}, \bibinfo
  {author} {\bibfnamefont {J.}~\bibnamefont {Arlt}}, \bibinfo {author}
  {\bibfnamefont {L.}~\bibnamefont {Santos}}, \bibinfo {author} {\bibfnamefont
  {G.}~\bibnamefont {T\'oth}}, \ and\ \bibinfo {author} {\bibfnamefont
  {C.}~\bibnamefont {Klempt}},\ }\href {\doibase
  10.1103/PhysRevLett.112.155304} {\bibfield  {journal} {\bibinfo  {journal}
  {Phys. Rev. Lett.}\ }\textbf {\bibinfo {volume} {112}},\ \bibinfo {pages}
  {155304} (\bibinfo {year} {2014})}\BibitemShut {NoStop}%
\bibitem [{\citenamefont {Chiuri}\ \emph {et~al.}(2012)\citenamefont {Chiuri},
  \citenamefont {Greganti}, \citenamefont {Paternostro}, \citenamefont
  {Vallone},\ and\ \citenamefont {Mataloni}}]{chiuri2012}%
  \BibitemOpen
  \bibfield  {author} {\bibinfo {author} {\bibfnamefont {A.}~\bibnamefont
  {Chiuri}}, \bibinfo {author} {\bibfnamefont {C.}~\bibnamefont {Greganti}},
  \bibinfo {author} {\bibfnamefont {M.}~\bibnamefont {Paternostro}}, \bibinfo
  {author} {\bibfnamefont {G.}~\bibnamefont {Vallone}}, \ and\ \bibinfo
  {author} {\bibfnamefont {P.}~\bibnamefont {Mataloni}},\ }\href {\doibase
  10.1103/PhysRevLett.109.173604} {\bibfield  {journal} {\bibinfo  {journal}
  {Phys. Rev. Lett.}\ }\textbf {\bibinfo {volume} {109}},\ \bibinfo {pages}
  {173604} (\bibinfo {year} {2012})}\BibitemShut {NoStop}%
\bibitem [{\citenamefont {Greenberger}\ \emph {et~al.}(1989)\citenamefont
  {Greenberger}, \citenamefont {Horne},\ and\ \citenamefont
  {Zeilinger}}]{greenberger1989}%
  \BibitemOpen
  \bibfield  {author} {\bibinfo {author} {\bibfnamefont {D.~M.}\ \bibnamefont
  {Greenberger}}, \bibinfo {author} {\bibfnamefont {M.~A.}\ \bibnamefont
  {Horne}}, \ and\ \bibinfo {author} {\bibfnamefont {A.}~\bibnamefont
  {Zeilinger}},\ }\href
  {http://inis.iaea.org/search/search.aspx?orig_q=RN:22064349} {\emph {\bibinfo
  {title} {Bell's theorem, quantum theory and conceptions of the universe}}}\
  (\bibinfo  {publisher} {Kluwer},\ \bibinfo {address} {Netherlands},\ \bibinfo
  {year} {1989})\BibitemShut {NoStop}%
\bibitem [{\citenamefont {Zeilinger}\ \emph {et~al.}(1992)\citenamefont
  {Zeilinger}, \citenamefont {Horne},\ and\ \citenamefont
  {Greenberger}}]{zeilinger1992}%
  \BibitemOpen
  \bibfield  {author} {\bibinfo {author} {\bibfnamefont {A.}~\bibnamefont
  {Zeilinger}}, \bibinfo {author} {\bibfnamefont {M.~A.}\ \bibnamefont
  {Horne}}, \ and\ \bibinfo {author} {\bibfnamefont {D.~M.}\ \bibnamefont
  {Greenberger}},\ }in\ \href {https://ntrs.nasa.gov/search.jsp?R=19920012809}
  {\emph {\bibinfo {booktitle} {Proceedings of Squeezed States and Quantum
  Uncertainty}}},\ \bibinfo {editor} {edited by\ \bibinfo {editor}
  {\bibfnamefont {D.}~\bibnamefont {Han}}, \bibinfo {editor} {\bibfnamefont
  {Y.~S.}\ \bibnamefont {Kim}}, \ and\ \bibinfo {editor} {\bibfnamefont
  {W.~W.}\ \bibnamefont {Zachary}}}\ (\bibinfo  {publisher} {NASA Conf. Publ.
  3135, 73},\ \bibinfo {year} {1992})\BibitemShut {NoStop}%
\end{thebibliography}%
\bibliographystyle{apsrev4-1}

\end{document}